\documentclass[11pt,a4 paper]{article}
\usepackage{graphicx,epsf,parskip,times,amsmath,lscape,multirow,colortbl,rotating, booktabs, subcaption,bm}
\usepackage[comma,sort&compress]{natbib}
\usepackage{xr}
\usepackage{url,booktabs}
\newcommand{\indep}{\perp \!\!\! \perp}

\begin{document}
\title{Prediction under interventions: evaluation of counterfactual performance using longitudinal observational data}

\author{RUTH H. KEOGH$^{1}$ \& NAN VAN GELOVEN$^{2}$ $^{\dag}$}

\date{\small {$^{1}$ \textit{Department of Medical Statistics, London School of Hygiene \& Tropical Medicine, London, UK}\\[2pt]
$^{2}$ \textit{Department of Biomedical Data Sciences, Leiden University Medical Center, Leiden, NL}
\\[2pt]
{ruth.keogh@lshtm.ac.uk, n.van\_geloven@lumc.nl}\\
{$^{\dag}$ The two authors contributed equally.}}}

\maketitle

\begin{abstract}
{Predictions under interventions are estimates of what a person’s risk of an outcome would be if they were to follow a particular treatment strategy, given their individual characteristics. Such predictions can give important input to medical decision making. However, evaluating predictive performance of interventional predictions is challenging. Standard ways of evaluating predictive performance do not apply when using observational data, because prediction under interventions involves obtaining predictions of the outcome under conditions that are different to those that are observed for a subset of individuals in the validation dataset. This work describes methods for evaluating counterfactual performance of predictions under interventions for time-to-event outcomes. This means we aim to assess how well predictions would match the validation data if all individuals had followed the treatment strategy under which predictions are made. We focus on counterfactual performance evaluation using longitudinal observational data, and under treatment strategies that involve sustaining a particular treatment regime over time. We introduce an estimation approach using artificial censoring and inverse probability weighting which involves creating a validation dataset that mimics the treatment strategy under which predictions are made.  We extend measures of calibration, discrimination (c-index and cumulative/dynamic AUCt) and overall prediction error (Brier score) to allow assessment of counterfactual performance. The methods are evaluated using a simulation study, including scenarios in which the methods should detect poor performance. Applying our methods in the context of liver transplantation shows that our procedure allows quantification of the performance of predictions supporting crucial decisions on organ allocation.}

\end{abstract}

\newpage
\section{Introduction}
\label{sec:intro}

Estimates of absolute risk of outcomes under different treatment choices conditional on patient characteristics are important for informing individual decisions in healthcare. This includes enabling patients to weigh their risks and benefits of different treatment options, and informing allocation of treatments that are subject to resource constraints, such as donor organs. Standard prediction models do not provide the necessary information as they target the observed outcome distribution in the population in which the prediction model was developed, typically including a mix of individuals with some who followed the treatment strategy of interest and others who did not \citep{van_geloven_prediction_2020}. The task of obtaining individualized estimates of risks under specified treatment strategies has been referred to as ‘causal prediction’, ‘counterfactual prediction’, and ‘prediction under hypothetical interventions’. In this paper we use the term ‘prediction under interventions’.

The fundamental challenge in developing and evaluating predictions under interventions is that after a patient receives a treatment and we observe their outcome, it is impossible to know what the counterfactual outcome would have been had they received an alternative treatment. Counterfactual predictions can be obtained from data collected in randomized controlled trials and some models for prediction under interventions have been developed and evaluated in this way \citep{nguyen_2020, Efthimiou_2023}. However, randomized trials are not designed for this purpose and may be limited by strict inclusion criteria, small sample size and short-term follow-up. Longitudinal observational data from sources such as electronic health records, cohort studies and patient registries, which provide rich data on large numbers of individuals, are often the main source for developing models for prediction under interventions. To address confounding, several causal inference methods that were originally proposed for estimating marginal treatment effects, have recently been adapted to develop prediction models under interventions conditional on baseline characteristics \citep{sperrin_using_2018, van_geloven_prediction_2020, lin_scoping_2021, dickerman:2022}. Methods that combine treatment effect estimates from trials with estimates of untreated risk from observational data have also been proposed \citep{xu_prediction_2021,sperrin_invited_2021}. 

The focus of this paper is performance evaluation, which is essential to inform adoption of interventional prediction models in practice. Evaluation of prediction models involves comparing estimated risks from the model with observed outcomes in a validation dataset. Treatment strategies targeted by an interventional prediction model will differ from those that are observed for a subset of the individuals in an observational validation dataset, meaning that there is no observed analogue of estimated risks for assessing predictive performance. This renders existing methods for evaluating predictive performance unusable in this setting. A recent review indicated that 0 out of 13 identified studies on prediction under interventions assessed performance and it has been described as the most pressing problem in this field \citep{lin_scoping_2021, sperrin_invited_2021}. Our goal is to assess counterfactual performance of an interventional prediction model. This means to assess how well the predictions would match the data if all individuals in the validation data had followed the treatment strategy under which predictions have been made. 

Prior works described approaches for counterfactual evaluation of interventional prediction models in the setting of a point treatment and binary outcome \citep{pajouheshnia_accounting_2017, coston_counterfactual_2020}. \cite{boyer_assessing_2023} recently formalized this and described extensions to longitudinal treatment strategies. These works did not cover time-to-event outcomes. An ad-hoc approach has been used in the time-to-event setting where estimated risks obtained under a given treatment strategy are compared with the observed outcomes in the subset of individuals who actually followed that treatment strategy \citep{sperrin_using_2018}. As we show later, this `subset approach' is prone to selection bias. 

In this paper we present the first general set of counterfactual performance measures for time-to-event outcomes. Our focus is on counterfactual performance evaluation using longitudinal observational data, and under treatment strategies that involve sustaining a particular treatment regime over time. The proposed approach is evaluated using a simulation study and we showcase its use by evaluating interventional prediction models in the context of liver organ allocation.

\section{Prediction under interventions}
\label{sec:est.dev}

We begin by defining the target of estimation for prediction under interventions, i.e. the estimand. Our focus is on the risk of an event up to a time horizon $\tau$ under specified longitudinal treatment strategies conditional on a set of predictors $X$. Earlier work targeted only untreated risk \citep{sperrin_using_2018,  van_geloven_prediction_2020, coston_counterfactual_2020, boyer_assessing_2023}. However, interest may lie in prediction under other longitudinal treatment strategies. Deterministic static strategies are arguably most relevant for informing individual decision making, and we let $\underline a_0$ denote a static deterministic treatment strategy from time zero onwards. In the later simulation and illustration we consider the strategies of never initiating treatment (\emph{never treated}, $\underline a_0=\mathbf 0$), and initiating treatment at time 0 and sustaining treatment thereafter (\emph{always treated}, $\underline a_0=\mathbf 1$). We let $T^{\underline a_0}$ denote the counterfactual event time under strategy $\underline a_0$. The estimand is the risk of the event occurring before time $\tau$ under this strategy given the predictors $X$ available at the time of making the prediction:
\begin{equation}
R^{\underline a_0} (\tau|X)=\Pr(T^{\underline a_0}\leq \tau|X).
\label{eq:risk}
\end{equation}
An overview of how $R^{\underline a_0} (\tau|X)$ can be estimated from longitudinal observational data using marginal structural model and cloning-censoring-weighting approaches is given in Supplementary Section \ref{sec:development}. For the remainder of the paper it is assumed that there exists a model that can be used to estimate $R^{\underline a_0} (\tau|X)$ for a new individual. 

\section{Counterfactual performance assessment}
\label{sec:eval}

\subsection{Validation data}

We assume that an external validation dataset with $n$ individuals is available for assessing counterfactual predictive performance of an interventional prediction model. The validation data should be from a population reflecting that in which the model is targeted for use \citep{sperrin_targeted_2022}. It is assumed to be observational and with a longitudinal structure in which each individual is observed at regular time points (e.g. study visits) $k=0,1,\ldots$ up to the event or censoring time, which is observed in continuous time. It must include the predictors $X$ plus any variables additionally needed, such that these in combination with $X$ form a valid adjustment set sufficient to control for confounding of the treatment-outcome association, including potential time-dependent confounding. 

We let $T$ and $C$ denote event and censoring times, measured relative to the time from which a prediction would be made. For individual $i$ the observed end of follow-up is $T_i^*=\mathrm{min}(T_i,C_i)$, and $D_i$ is the event indicator. Time-dependent covariates $L_k$ and treatment status $A_k$ are recorded at each visit. We assume all individuals are untreated before time 0 ($A_{0^-}=0$), but may follow any treatment pattern thereafter. The structure of the validation dataset is illustrated in the directed acyclic graph (DAG) in Figure \ref{fig:dag}, which depicts presence of time-dependent confounding by $L_k$. The predictors $X$ may include all or a subset of the baseline confounders $L_0$, denoted $L_0^*$, in addition to variables $P$ that predict the outcome but do not affect treatment status, i.e. $X=\{L_0^*, P\}$. 

The interventional prediction model is applied for each individual $i=1,\ldots,n$ in the validation data to obtain $\hat R^{\underline a_0} (\tau|X_i)$, the estimate of risk up to time $\tau$ if they were to follow treatment strategy $\underline a_0$.

\subsection{Artificial censoring and inverse probability weighting}
\label{subsec:eval.ipcw}

Our proposed approach uses the observed validation data to generate modified validation datasets that emulate scenarios in which every subject had followed each treatment strategy of interest. The first step involves applying artificial censoring in the validation data such that an individual's follow-up is censored when they deviate from the strategy of interest, which could be at time 0. This is illustrated in Supplementary Figure \ref{fig:method.suppl} for the always treated ($\underline a_0=\mathbf 1$) and never treated ($\underline a_0=\mathbf 0$) strategies, but can be applied for any strategy for which predictions can be obtained provided there is a sufficiently large subgroup of individuals in the validation data who follow that strategy. The modified validation dataset restricted to follow-up under strategy $\underline a_0$ is denoted $V^{\underline a_0}$. 
The second step involves weighting the individuals in $V^{\underline a_0}$ in such a way that it represents the population as if all individuals in the validation data had followed treatment strategy $\underline a_0$. The weight at time $t$ is the inverse of the probability of not having been artificially censored up to that time, conditional on the covariate history required to control confounding, which in our DAG is $\bar{L}_t$.
 
 In $V^{\underline a_0}$, let $C_{\underline a_0}$ denote the artificial censoring time and $\tilde{T}_{\underline a_0}=\min(T^*,C_{\underline a_0})$ the observed event time after artificial censoring with event indicator $\tilde{D}_{\underline a_0}= I(T^*<C_{\underline a_0})D$. The inverse probability of artificial censoring weights (IPACW) are:
   \begin{equation}
  \label{eqn:weights}
     G_{\underline a_0}^{-1}(t|L)=\prod_{s=0}^{\lfloor t \rfloor}\Pr(C_{\underline a_0}>s|C_{\underline a_0}>s-1,\bar L_s)^{-1} =\prod_{s=0}^{\lfloor t \rfloor}\frac{1}{\Pr(A_s=a_s|\bar A_{s-1}=\bar a_{s-1},\bar L_s)} 
 \end{equation}
 The models used to derive the weights should be fitted in the validation data and not confused with any weight model fitted in the development data, as treatment assignment may be different in the development and validation datasets. 

Some estimators for evaluating predictive performance in the (non-interventional) prediction setting apply inverse probability of censoring weights to account for standard right censoring such as loss-to-follow-up or end of study ($C$ above). We assume the standard censoring times do not depend on covariates (with $G_c(t)=P(C>t)$) but this could be extended to covariate-dependent censoring. Assuming independence between the artificial and standard censoring processes, we define the combined weight:
\begin{equation}
    \label{eqn:weights.ac}
G_{\underline a_0 c}^{-1}(t|L)=G_{\underline a_0}^{-1}(t|L)\times G_c^{-1}(t).
\end{equation}

The validity of our approach, and of the estimates of counterfactual performance described below, relies on the causal assumptions of 
\begin{itemize}
    \item {conditional sequential exchangeability: $T^{\bar{A}_{k-1},\underline{a}_{k}}\indep A_k|\bar{L}_{k},P,\bar{A}_{k-1},T\geq k$}
    \item {consistency: $T=T^{\underline{a}_{0}}$ if $\underline{A}_0=\underline a_0$}
    \item {positivity: $0<\Pr(A_k=a_k|\bar L_k,P,\bar A_{k-1}=\bar a_{k-1})<1$}
\end{itemize}
for all $k$.
It also relies on on correct specification of the models used to estimate the weights.

\section{Counterfactual performance measures
}
\label{sec:calib}

We describe counterfactual measures of calibration, discrimination, and overall prediction error for validation of predictions under interventions. An overview of these measures in the standard prediction setting for time-to-event outcomes is given by \cite{mclernon_assessing_2022}. We specify estimands for each measure, and describe estimators, which extend previously proposed estimators for the standard prediction setting, in particular by adding weights that depend on time-dependent covariates.

\subsection{Counterfactual calibration}
\label{subsec:calib}

Calibration assessment focuses on how close estimates of risk by a particular time horizon $\tau$ from a prediction model are to the true underlying risks. For assessment of counterfactual performance under strategy $\underline a_0$, mean calibration compares the average estimated risk, $\bar {R}^{\underline a_0}(\tau)=\frac{1}{n}\sum_{i=1}^{n}\hat R^{\underline a_0}(\tau|X_i)$, with the counterfactual outcome proportions by time $\tau$ under strategy $\underline a_0$, $R_{\mathrm{Obs}}^{\underline a_0}(\tau)$. An estimate of $R_{\mathrm{Obs}}^{\underline a_0}(\tau)$, denoted $\hat R_{\mathrm{Obs}}^{\underline a_0}(\tau)$, can be obtained by applying a weighted Kaplan-Meier analysis to $V^{\underline a_0}$, with weights $G_{\underline a_0}^{-1}(\tau|L)$. The estimates can be compared using the observed versus expected (OE) ratio $\hat R_{\mathrm{Obs}}^{\underline a_0}(\tau)/\bar R^{\underline a_0}(\tau)$.


A stronger assessment of calibration is how well estimated risks from the prediction model agree with the observed outcome proportions across the range of risk. To extend this to counterfactual performance assessment we divide the predictions $\hat R^{\underline a_0}(\tau|X_i)$ ($i=1,\ldots,n$) into $G$ equal sized groups (e.g. $G=10$), with the mean prediction in group $g$ denoted $\bar {R}_g^{\underline a_0}(\tau)$. We then estimate the counterfactual outcome proportions in each group $g=1,\ldots,G$, denoted $\hat R_{\mathrm{Obs},g}^{\underline a_0}(\tau)$, using a weighted Kaplan-Meier analysis for group $g$ in $V^{\underline a_0}$, with the same weights as above. The $\hat R_{\mathrm{Obs},g}^{\underline a_0}(\tau)$ ($g=1,\ldots,G$) can then be plotted against $\bar {R}_g^{\underline a_0}(\tau)$ ($g=1,\ldots,G$) to visually assess calibration.

\subsection{Counterfactual discrimination}
\label{subsec:disc}

Concordance statistics for time-to-event outcomes, such as the c-index and the time-dependent area under the ROC curve (AUCt), compare pairs of individuals and evaluate whether the individual with the shorter survival time was assigned the higher risk by the model. In the presence of censoring, pairs in which one individual is censored before the other has an observed event are not `comparable', and in a standard prediction context weighted estimators for the c-index and AUCt have been derived to address this \citep{uno_evaluating_2007, uno_c-statistics_2011, gerds_estimating_2013, blanche_review_2013}. 
We extend these to allow the weights to incorporate time-dependent covariates.

We define the counterfactual c-index up to time horizon $\tau$ under treatment strategy $\underline a_0$ as 
$C^{\underline a_0}(\tau)=\Pr(\hat R^{\underline a_0}_i(\tau|X_i) > \hat R^{\underline a_0}_j(\tau|X_j)| T^{\underline a_0}_i<T^{\underline a_0}_j, T^{\underline a_0}_i\leq \tau )$, 
with $\hat R^{\underline a_0}_i(\tau|X_i)$ and $\hat R^{\underline a_0}_j(\tau|X_j)$ the predictions for a pair of individuals $i$ and $j$.

To estimate $C^{\underline a_0}(\tau)$, we use comparable pairs of individuals in $V^{\underline a_0}$. Time-dependent weights $G_{\underline a_0c}^{-1}(t|L)$ (equation \ref{eqn:weights.ac}) are applied to account for both artificial censoring and standard censoring. We propose the following weighted estimator:
\begin{equation}
\hat{C}^{\underline a_0}(\tau)=\frac{\sum_{i=1}^{n} \sum_{j=1}^{n} I(\hat R^{\underline a_0}_i(\tau|X_i)>\hat R^{\underline a_0}_j(\tau|X_j))\mathrm{comp}^{(1)}_{\underline a_0,ij}(\tau)\hat{W}^{(1)}_{\underline a_0,ij} }
{\sum_{i=1}^{n} \sum_{j=1}^{n} \mathrm{comp}^{(1)}_{\underline a_0,ij}(\tau)\hat{W}^{(1)}_{\underline a_0,ij}}
\end{equation}
where $\mathrm{comp}^{(1)}_{\underline a_0,ij}(\tau)=I(\tilde{T}_{\underline a_0i}<\tilde{T}_{\underline a_0j}, \tilde{T}_{\underline a_0i}\leq \tau, \tilde{D}_{\underline a_0i}=1)$ indicates whether the pair $(i,j)$ is comparable up to time $\tau$ in $V^{\underline a_0}$, and $\hat{W}^{(1)}_{\underline a_0,ij}=\hat{G}_{\underline a_0c}^{-1}(\tilde{T}^{-}_{\underline a_0i}|{L}_{i})\hat{G}_{\underline a_0c}^{-1}(\tilde{T}_{\underline a_0i}|{L}_{j})$ is the weight of the pair, where $\hat{G}_{\underline a_0c}(\tilde{T}^{-}_{\underline a_0i}$) is the left hand limit of $\hat{G}_{\underline a_0c}(\tilde{T}_{\underline a_0i})$.   

Corresponding results for the cumulative/dynamic AUCt for prediction under interventions are in Supplementary Section \ref{sec:auc.suppl}. 

Identification of these discrimination indices requires extension of the positivity assumption to ensure a nonzero number of comparable pairs under the treatment strategies of interest \citep{boyer_assessing_2023}.

\subsection{Counterfactual Brier score}
\label{subsec:brier}

The Brier score measures overall predictive performance and extends mean squared error to the time-to-event setting \citep{graf_assessment_1999}. 
The estimand for the Brier score for counterfactual performance under treatment strategy $\underline a_0$ is defined as $BS^{\underline a_0}(t)=E[(I(T^{\underline a_0}\leq t)-\hat R^{\underline a_0}(t|X))^2]$. 
 
In $V^{\underline a_0}$, only observations not censored before $t$ contribute to the calculation of $BS^{\underline a_0}(t)$. The proposed estimator weighs these observations by the inverse of their probability of remaining uncensored: 
\begin{equation}
\hat{BS}^{\underline a_0}(t)=\frac{1}{n}\sum_{i=1}^{n} ((I(\tilde{T}_{\underline a_0i}\leq t)-\hat R^{\underline a_0}_i(t|X_i))^2W^{(2)}_{\underline a_0i}),
\label{eq:brier}
\end{equation}
with $W^{(2)}_{\underline a_0i}=\frac{I(\tilde{T}_{\underline a_0i}\leq t, \tilde{D}_{\underline a_0i}=1)}{\hat{G}_{\underline a_0c}(\tilde{T}_{\underline a_0i}|{L}_i)}+\frac{I(\tilde{T}_{\underline a_0i}>t)}{\hat{G}_{\underline a_0c}(t|{L}_i)}$, representing weights for individuals whose event was observed before $t$ and individuals observed to stay event-free by time $t$ in $V^{\underline a_0}$. This extends a weighted estimator for baseline covariate-dependent censoring \citep{gerds_consistent_2006}. 

The scaled Brier score under treatment strategy $\underline a_0$ is
$BS^{\underline a_0}_{\mathrm{scaled}}=1-BS^{\underline a_0}(t)/BS^{\underline a_0}_0(t)$,
where $BS^{\underline a_0}_0(t)$ is the Brier score of the null model. $BS^{\underline a_0}_0(t)$ can be estimated using (\ref{eq:brier}), with $\hat R^{\underline a_0}(t|X_i)$ replaced by the risk up to time $t$, which is estimated by a weighted sum of event indicators in $V^{\underline a_0}$ at time $t$, with $\hat G^{-1}_{\underline a_0c}(t)$ as weights.

\section{Simulation study}
\label{sec:sim}

\subsection{Simulation plan}
\label{sec:sim.plan}

We evaluate the performance of the proposed measures of counterfactual predictive performance 
in a simulation study. In each simulation run, we first generate a development dataset and use this to derive a model for prediction under \emph{never treated} and \emph{always treated} strategies. Next, we generate a longitudinal observational validation dataset (Supplementary Figure \ref{fig:dag.suppl}), and obtain predictions under the \emph{never treated} and \emph{always treated} strategies using the development model. We estimate predictive performance of the interventional predictions in the validation data, applying both the proposed approach to assess counterfactual performance and the ad-hoc subset approach. These estimates are compared against predictive performance derived from two `perfect' validation datasets, one for each of the \emph{never treated} and \emph{always treated} strategies, generated in such a way that everyone followed the treatment strategy of interest. 

Three main scenarios are considered. In Scenario 1, the development and validation datasets are generated under the same model. 
In Scenario 2, the development dataset has a higher baseline hazard than the validation dataset, but the form of the hazard model is otherwise the same. In Scenario 3, the development and validation datasets are generated under the same model, but the predictions in the validation data are obtained using an error prone version of $L_0$, denoted $L_0^*$. Scenarios 2 and 3 mimic settings where we expect poor counterfactual performance. 
In the three main simulation scenarios the assumptions of consistency, positivity, conditional exchangeability and correct specification of the weights model, hold. In three additional scenarios, we examine where and how our method breaks down when one of these assumptions is violated.
For all scenarios we consider data generating mechanisms using an additive hazards model and using a proportional hazards model.
Full details of the simulation plan are given in Supplementary Section 
\ref{sec:sim.plan.suppl} and descriptives of the resulting datasets in Supplementary Section \ref{sec:sim.results.suppl}. 

R code for replicating the simulation is provided at \url{https://github.com/survival-lumc/Validation_Under_Interventions}.

\subsection{Simulation results}
\label{sec:sim.results}

\begin{figure}
	\centering
		\includegraphics[scale=0.95]{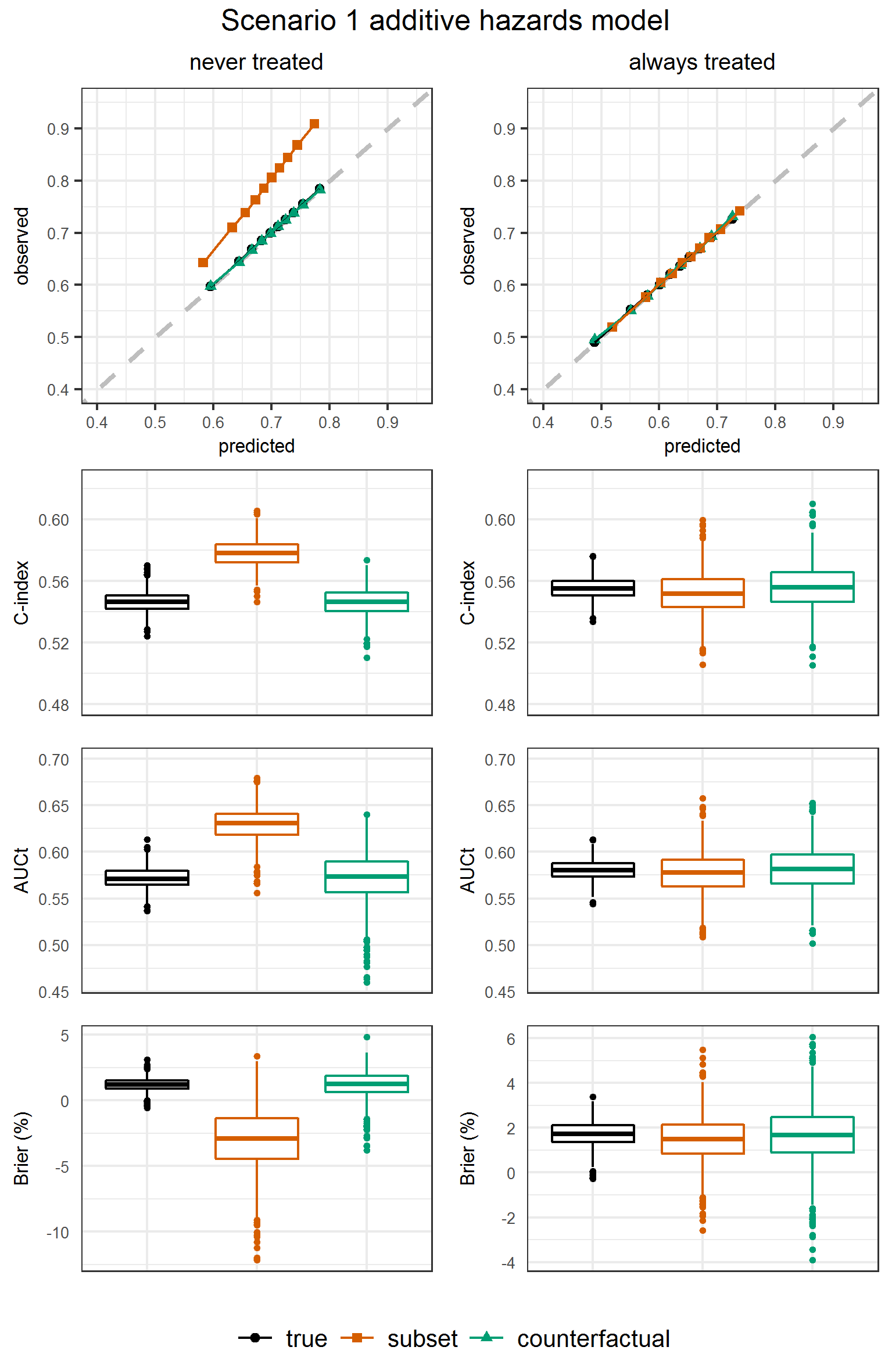}
	    \caption{Simulation results: additive hazards model Scenario 1. Left panel:  for the \emph{never treated} strategy. Right panel: for the \emph{always treated} strategy. Performance measures were obtained from the perfect validation data (black dots) and estimated from the observational validation data using the subset approach (orange squares) and using the proposed artificial censoring + inverse probability weighted estimators of counterfactual performance (green triangles). Top row: calibration plot showing observed outcome proportions against mean estimated risks by time 5 within tenths of the estimated risks. Second row: c-index truncated at time 5. Third row: cumulative/dynamic area under the ROC curve at time 5. Bottom row: scaled Brier score at time 5.}\label{fig:sim_addhaz_scenario1}
\end{figure}

\begin{figure}
	\centering
		\includegraphics[scale=.95]{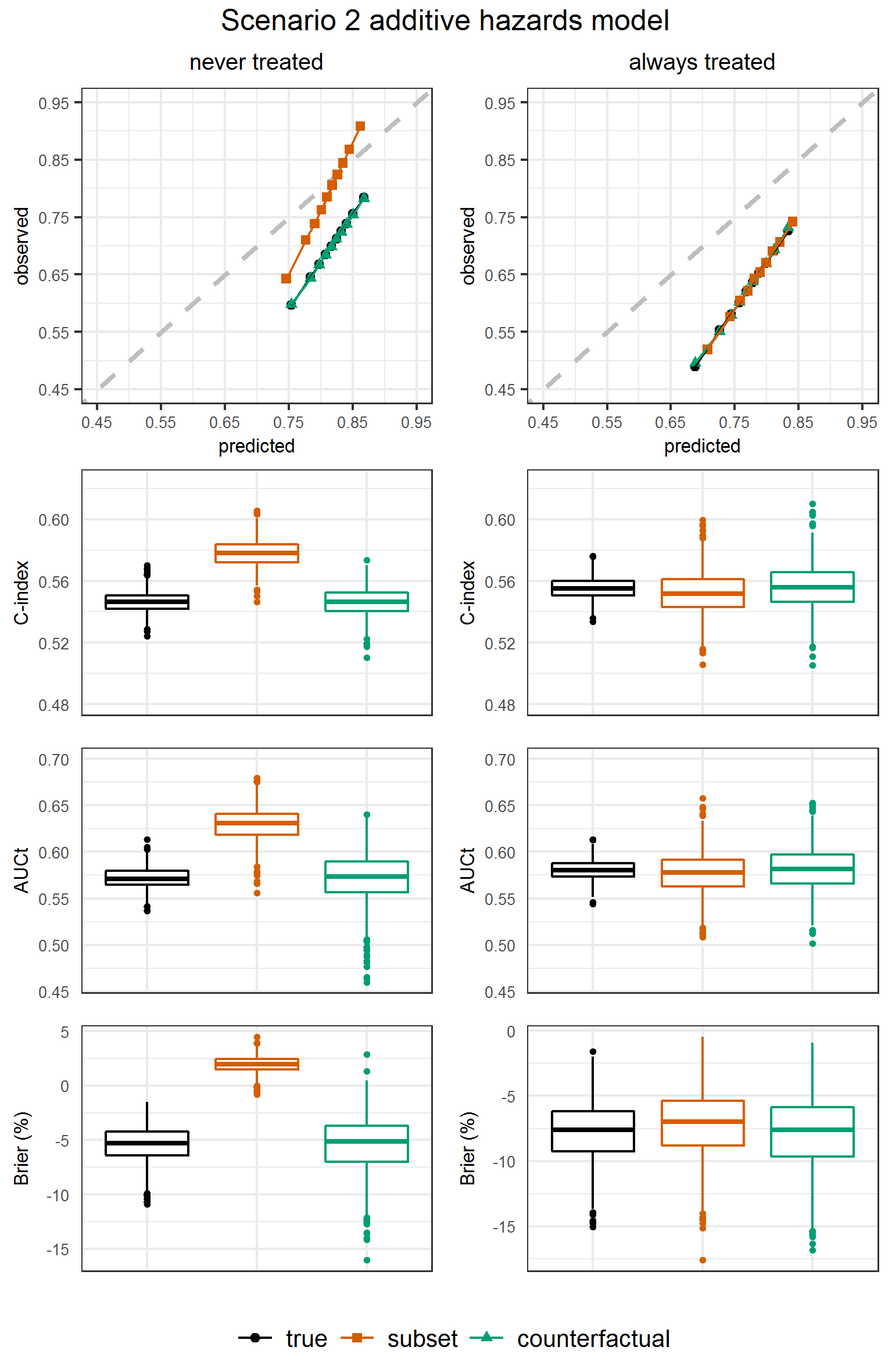}
	    \caption{Simulation results: additive hazards model Scenario 2. Left panel:  for the \emph{never treated} strategy. Right panel: for the \emph{always treated} strategy. Performance measures were obtained from the perfect validation data (black dots) and estimated from the observational validation data using the subset approach (orange squares) and using the proposed artificial censoring + inverse probability weighted estimators of counterfactual performance (green triangles). Top row: calibration plot showing observed outcome proportions against mean estimated risks by time 5 within tenths of the estimated risks. Second row: c-index truncated at time 5. Third row: cumulative/dynamic area under the ROC curve at time 5. Bottom row: scaled Brier score at time 5.}\label{fig:sim_addhaz_scenario2}
\end{figure}

\begin{figure}
	\centering
		\includegraphics[scale=0.95]{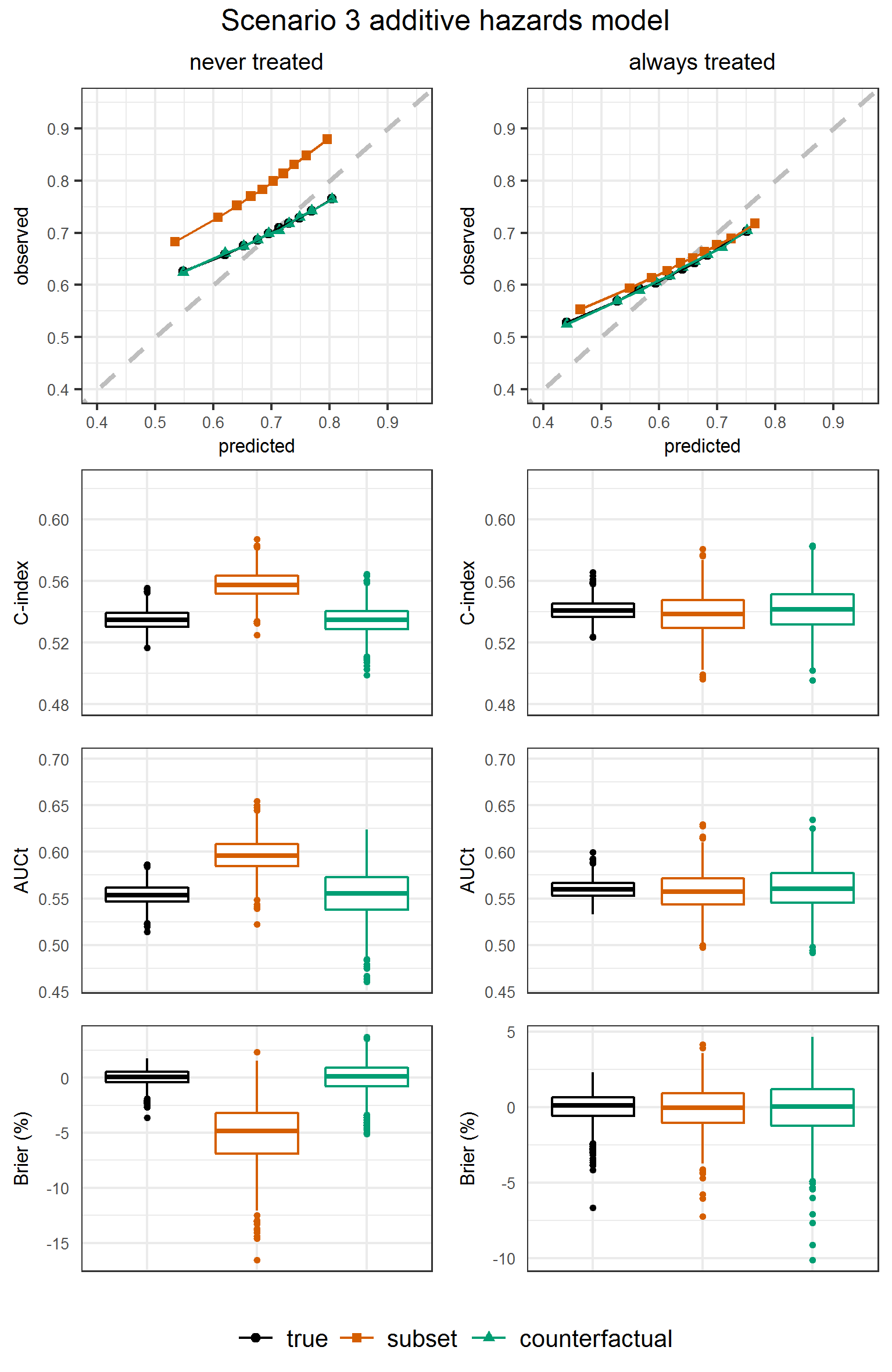}
	    \caption{Simulation results: additive hazards model Scenario 3. Left panel:  for the \emph{never treated} strategy. Right panel: for the \emph{always treated} strategy. Performance measures were obtained from the perfect validation data (black dots) and estimated from the observational validation data using the subset approach (orange squares) and using the proposed artificial censoring + inverse probability weighted estimators of counterfactual performance (green triangles). Top row: calibration plot showing observed outcome proportions against mean estimated risks by time 5 within tenths of the estimated risks. Second row: c-index truncated at time 5. Third row: cumulative/dynamic area under the ROC curve at time 5. Bottom row: scaled Brier score at time 5.}\label{fig:sim_addhaz_scenario3}
\end{figure}

Figures \ref{fig:sim_addhaz_scenario1}, \ref{fig:sim_addhaz_scenario2} and \ref{fig:sim_addhaz_scenario3} show results for calibration, c-index, AUCt and the scaled Brier score in Scenarios 1-3, with data generated and analysed using an additive hazards model. Corresponding numerical results 
are presented in Supplementary Tables \ref{tab:addhaz_sc1}, \ref{tab:addhaz_sc2} and \ref{tab:addhaz_sc3}, where we also show the ratio of observed to estimated risks by time 5 (OE ratios). Results from the other scenarios and from scenarios obtained from using a proportional hazards model to generate and analyse the data are presented in Supplementary Section \ref{sec:sim.results.suppl}. 

In Scenario 1, where the developed prediction model is correctly specified and the development and validation data are generated from the same distributions, the calibration plot (top panel Figure \ref{fig:sim_addhaz_scenario1}) shows that our proposed method for counterfactual performance evaluation correctly assesses that estimated and observed outcome proportions lie on the diagonal (perfect calibration), with OE ratios on average estimated very close to one (Supplementary Table \ref{tab:addhaz_sc1}). The subset approach wrongly suggested miscalibration for the predictions under the \emph{never treated} strategy. 
The counterfactual evaluation of discrimination resulted in correct estimates of c-index and AUCt whereas the subset approach overestimated these indices 
for the \emph{never treated} strategy (middle panels Figure \ref{fig:sim_addhaz_scenario1}, Supplementary Table \ref{tab:addhaz_sc1}). The subset approach estimated a negative scaled Brier score on average in the \emph{never treated} strategy (lower panel Figure \ref{fig:sim_addhaz_scenario1}) suggesting that the developed model was no better than a null model assigning the average risk to all subjects in the validation data, contrary to the true positive scaled Brier score. The proposed methods for counterfactual performance evaluation also showed unbiased results for the \emph{always treated} strategy. 
Discussion of the size and direction of the biases when using the subset approach is presented in Supplementary Section \ref{sec:sim.results.suppl}. 

In Scenario 2, the generated risks were based on a higher baseline hazard in the development data compared to the validation data. The 
 counterfactual measures of calibration correctly detected the resulting overestimation of estimated risks compared to observed outcome proportions 
 with the calibration curves lying below the diagonal (Figure \ref{fig:sim_addhaz_scenario2}), the OE ratios being below one and negative scaled Brier scores for both treatment strategies (Supplementary Table \ref{tab:addhaz_sc2}). The subset approach gave estimated OE ratios that are too high and for the \emph{never treated} strategy it incorrectly suggested a positive scaled Brier score. Discrimination results are unaffected by changes in baseline hazard only, so the results for c-index and AUCt are equal to those in Scenario 1. 

In Scenario 3 we used an error-prone version of $L_0$ when obtaining risk estimates in the validation data. The resulting 
lower levels of c-index, AUCt and scaled Brier score 
were correctly picked up by the counterfactual performance measures (Figure \ref{fig:sim_addhaz_scenario3}, Supplementary Table \ref{tab:addhaz_sc3}). The counterfactual calibration plot correctly showed that using the error prone $L_0$ results in estimated risks that are too extreme, such that the very high and very low estimated risks correspond to observed outcome proportions that are closer to the average (top panel Figure \ref{fig:sim_addhaz_scenario3}). Averaged over all patients the risks are still well calibrated, with the OE ratios around one. 
The subset approach again overestimates OE ratios and discrimination indices in the \emph{never treated} strategy. 

Results for the other performance measures in scenarios 1-3 are presented in Supplementary Figures \ref{fig:sim.addhaz.scenario1.appendix}, \ref{fig:sim.addhaz.scenario2.appendix} and \ref{fig:sim.addhaz.scenario3.appendix} and confirm the above observations. 

As expected, the scenarios with deliberately introduced violations of causal assumptions in the validation data showed bias in the estimates of counterfactual performance. Bias varied across the different degrees of violations. Numerically, bias was modest in the counterfactual discrimination measures and was more pronounced in the estimates of the OE ratio for calibration in some scenarios. Despite the violations, bias of the counterfactual performance metrics was smaller than that of the naive subset approach in 58/64 (90\%) of times in the additive hazards model Scenarios (Supplementary Table \ref{tab:addhaz_sc456}). 


\section{Application to liver transplantation}
\label{sec:example}


We illustrate our methods with an application to liver transplantation, using data from the Scientific Registry of Transplant Recipients (SRTR). The SRTR data system includes data on all donors, wait-listed candidates, and transplant recipients in the US, submitted by the members of the Organ Procurement and Transplantation Network (OPTN). The
Health Resources and Services Administration (HRSA), U.S. Department of Health and Human
Services provides oversight to the activities of the OPTN and SRTR contractors.

 We consider a composite outcome of death or removal from the transplant waitlist due to worsening health status. Our aim is to estimate for all wait-listed individuals at any given time their risk of the composite outcome up to 3 years under two intervention strategies: (1) receiving a liver transplant at that time; (2) not receiving a transplant at that time or in the future. The predictions under each intervention are to be conditional on the most recent measurements of individual characteristics. At the moment a new donor organ becomes available, such predictions would enable decision-makers to weigh the estimated risks of all wait-listed candidates under both strategies, which could inform organ allocation. 


We used data on 43190 individuals who joined the liver transplant waitlist between 1 January 2014 and 30 April 2019. 
Information recorded includes date of receiving a transplant, date of death, and date of and reason for removal from the waitlist, 
alongside longitudinal measurements of biomarkers, complications and comorbidities. We created an analysis dataset that combines: (1) a dataset of individuals followed-up from transplant onwards; (2) datasets starting at a series of landmark times (from the time of joining the waitlist) restricted to individuals who remain on the waitlist and are untransplanted at the landmark time. The combined dataset was divided randomly into a 70\% sample used for model development and a 30\% sample used for the validation. Prediction models under the two interventions were fitted using the development dataset, applying an extension of methods used previously to estimate the effects of transplant to the prediction under interventions setting \citep{gong_estimating_2017,strohmaier_survival_2022}. The validation methods proposed above were then used. 
Details about data set-up, model development, and how the validation methods were applied are given in Supplementary Section \ref{sec:application.suppl}. 

\begin{figure}
    \centering
    \includegraphics[scale=0.8]{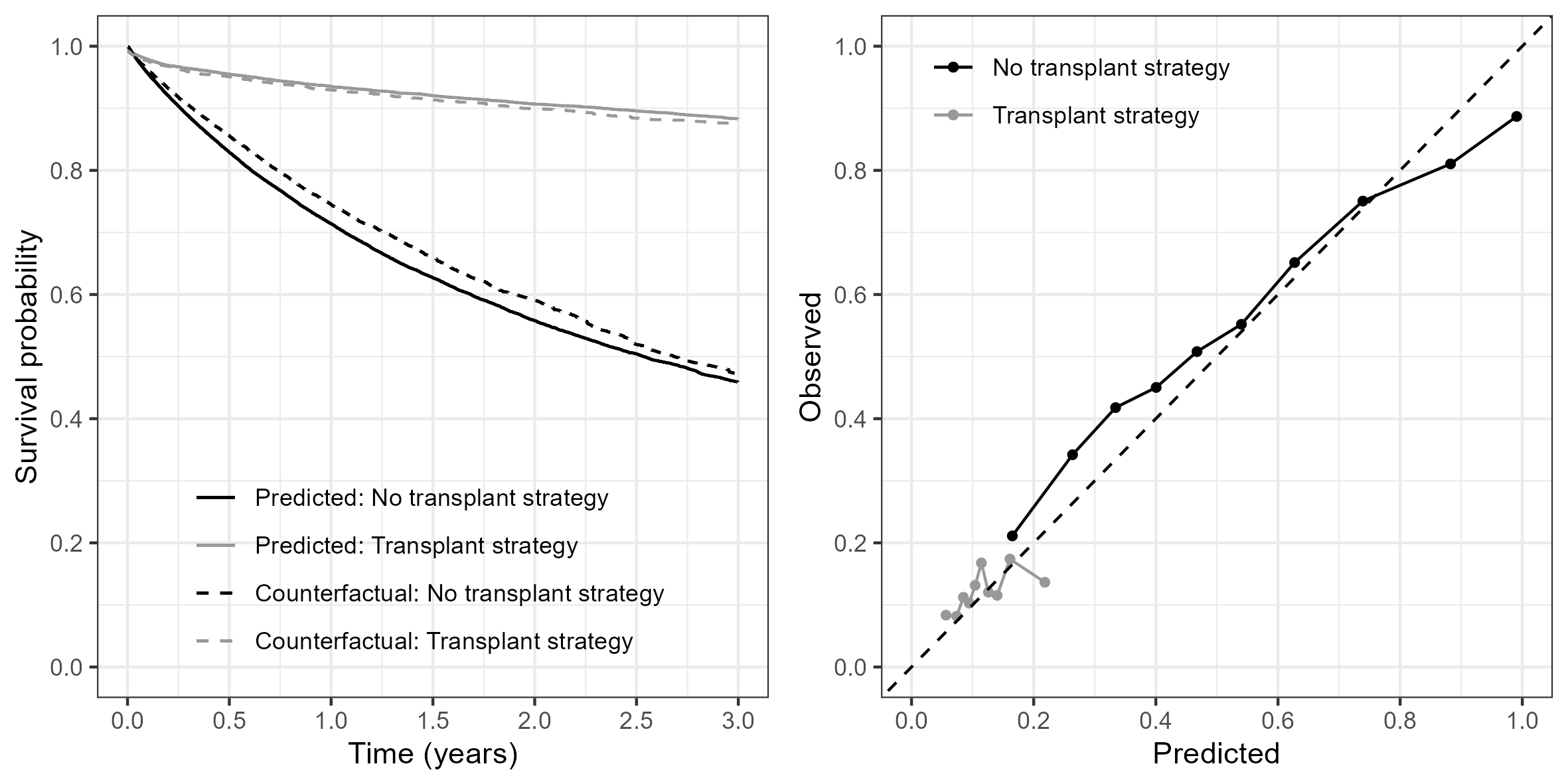}
    \caption{Liver transplant application: Calibration of estimated risks under the \emph{no transplant} and \emph{transplant} strategies. Left: Plot showing the mean estimated survival curves up to three years under the two transplant strategies (solid lines), and the corresponding counterfactual survival curves (dashed lines), obtained using the artificial censoring plus inverse probability weighting approach. Right: Plot of counterfactual outcome proportions by 3 years against mean estimated risk by 3 years within 10 equal-sized groups of estimated risk under the two transplant strategies, showing the line of equality (dashed line).}
    \label{fig:liver.calibration}
\end{figure}

Figure \ref{fig:liver.calibration} shows calibration plots. Mean predicted survival curves differ substantially between the two intervention strategies, with mean estimated risk by 3 years being 54.1\% under the \emph{no transplant} strategy and 11.7\% under the \emph{transplant} strategy. The `observed' overall survival curves are close to the mean predicted curve, though the predicted curve under the \emph{no transplant} strategy is slightly too low. Mean calibration is better for the \emph{transplant} strategy. The comparison of mean risk estimates for $\tau=3$ years within ten groups  with the `observed' outcome proportions within each group suggests some overfitting for the \emph{no transplant} strategy: low risks tending to be underestimated and high risks overestimated. A similar pattern is seen for the \emph{transplant} strategy, for which we see a narrow range of predicted risks. 
Table \ref{tab:liver} summarises the other validation metrics. The c-index up to 3 years is 0.749 for prediction under the \emph{no transplant} strategy and 0.561 for prediction under the \emph{transplant} strategy. The AUCt's are 0.781 and 0.552, and scaled Brier scores are 66.8\% and 12.0\%. 

The counterfactual validation approach allowed us to assess all relevant aspects of model performance. The results could inform improvements to these models or comparison with alternative ones. 
The subset approach would have given an incorrect impression of the performance of the interventional prediction models (Supplementary Table \ref{tab:liver:subset} and Supplementary Figure \ref{fig:liver.calibration.subset}). 

\section{Discussion}
\label{sec:discussion}

In this paper we proposed a new approach for counterfactual validation of predictions under interventions for time-to-event outcomes. We allow use of longitudinal observational data with time-dependent confounding, and for interventions that involve sustaining a treatment over time. Our proposed approach is based on creating modified validation datasets that emulate scenarios in which every individual follows each treatment strategy of interest, through the use of artificial censoring and inverse probability weighting. We have provided the first general set of measures of counterfactual predictive performance for time-to-event outcomes, including measures of calibration, discrimination, and overall prediction error. Our simulation study showed that the proposed measures 
correctly capture true predictive performance, including detecting poor performance when they should. An application in the context of liver transplantation showed that our procedure allows quantification of the performance of predictions supporting crucial decisions on organ allocation.

Our validation approach relies on correct specification of the models used to generate the inverse probability weights, and on the assumptions of consistency, positivity and conditional exchangeability. Simulation scenarios with violation of these assumptions showed the bias introduced when they do not hold. To avoid positivity violations and in general large uncertainty in estimates, it is advised to only consider treatment strategies that are followed by a sufficiently large subgroup in the validation data. Pragmatic descriptions of treatment strategies may help in this regard. In general, the censoring and weighting approach without a structural model is less efficient compared to approaches where treatment effects are modelled \citep{hernan_comparison_2006}. Alternative approaches that employ a type of outcome modelling in the validation step have been proposed for counterfactual evaluation of binary outcomes predictions \citep{coston_counterfactual_2020, boyer_assessing_2023}. Extensions of our approach that allow relaxation of modelling assumptions, such as doubly robust approaches, are a priority for future work. 
In the longitudinal time-to-event setting, such approaches would require modelling of longitudinal data over time, in line with the g-formula approach to predictions under interventions used by \cite{dickerman:2022}.

 We studied a setting in which the validation data have longitudinal information on treatment use and on time-fixed and time-dependent confounders, alongside the event or censoring time, but our approach can also be used in simpler settings, for example with point treatments or only time-fixed confounders. 
Our validation approach is also straightforward to extend to more complex treatment strategies, such as dynamic regimes. It is important to emphasise that validation of interventional predictions requires that the validation data includes not only the predictors, but also any additional variables required to control confounding, and information on starting and stopping of treatment.

In the simulation study we imagined an external validation dataset, but 
many prediction studies rely initially on internal validation. In the application we used a split sample approach. 
Further work is needed to demonstrate how our validation methods can be extended for use with cross-validation and bootstrapping. 
For example, in cross-validation, if each fold has limited sample size then stable estimation of the weights could be challenging. In addition, methods for establishing sample size requirements are an area for future work, as are methods for model updating such as using recalibration. 

While methods for development of interventional prediction models have been increasing, there has been very little focus on the validation of the resulting predictions. Assessment of counterfactual predictive performance is a pivotal step towards implementation of interventional predictions models and may provide a novel instrument for model selection and tuning. 





\section*{Software}
\label{sec:software}

R code for replicating the simulation is provided at \url{https://github.com/survival-lumc/Validation_Under_Interventions}.

\section*{Acknowledgments}

The data reported here have been supplied by the Hennepin Healthcare Research Institute (HHRI) as the contractor for the Scientific Registry of Transplant Recipients (SRTR). The interpretation and reporting of these data are the responsibility of the author(s) and in no way should be seen as an official policy of or interpretation by the SRTR or the U.S. Government. The authors thank Ilaria Prosepe (Leiden University Medical Center, NL) for advice on the application. RHK was funded by UK Research and Innovation (Future Leaders Fellowship MR/S017968/1).

\clearpage

\bibliographystyle{biorefs}
\bibliography{biostatistics_references}

\clearpage

\begin{table}
\caption{Liver transplant application: Evaluation of counterfactual performance.}
\label{tab:liver}
\begin{center}
\begin{tabular}{lcc}
\hline
&\multicolumn{2}{c}{Strategy}\\
\cline{2-3} 
&No transplant&Transplant\\
\hline
  Calibration: OE ratio based on risk by 3 years  &0.983 &1.060\\
&&\\
  Discrimination: C-index up to 3 years&0.749&0.561\\
  Discrimination: AUCt at 3 years &0.781&0.552\\
  &&\\
  Prediction error: scaled Brier score (\%) at 3 years &66.8&12.0\\
  \hline
  {\small OE ratio: observed versus expected ratio.}\\
    {\small AUCt: cumulative/dynamic area under the ROC curve.}\\
\end{tabular}
\end{center}
\end{table}

\clearpage

\begin{figure}
	\centering
		\includegraphics[scale=1.2]{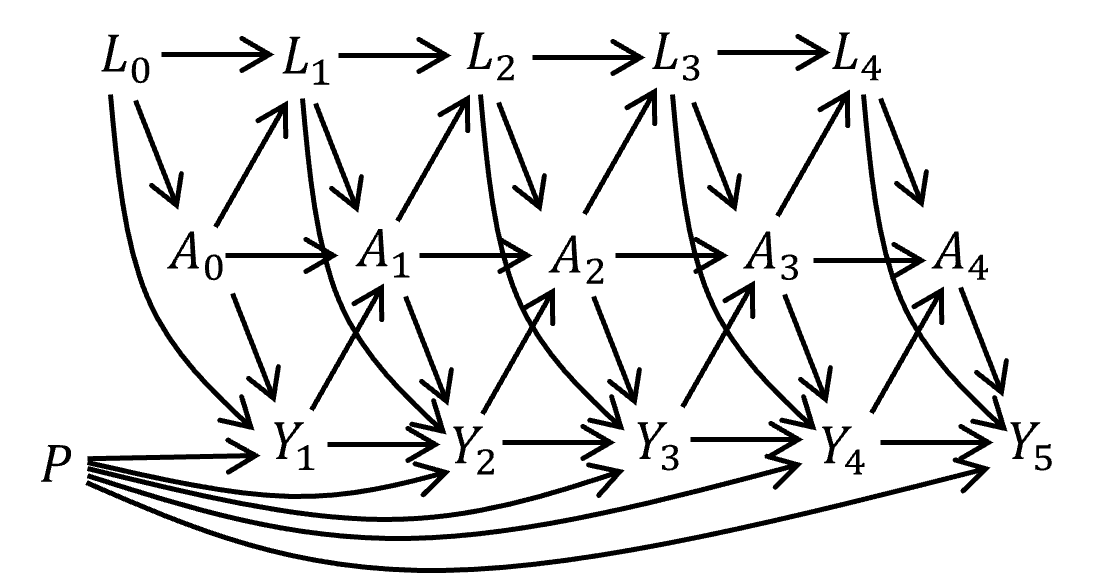}
			\caption{Directed acyclic graph (DAG) illustrating relationships between treatment $A$, time-dependent covariates $L$, baseline prognostic variables $P$, and discrete time outcome $Y$. The DAG is illustrated for a discrete-time setting where $Y_k=I(k-1 \leq T< k)$ is an indicator of whether the event occurs between visits $k-1$ and $k$. If the DAG is extended by adding a series of small time intervals between each visit, at which events are observed, then we approach the continuous time setting. The covariates $L_k$ are time-dependent confounders as they inform treatment initiation or continuation, are predictive of the outcome, and are also affected by past treatment.}\label{fig:dag}
\end{figure}


\clearpage

\setcounter{section}{0}
\renewcommand{\thesection}{S\arabic{section}}

\setcounter{table}{0}
\renewcommand{\thetable}{S\arabic{table}}

\setcounter{figure}{0}

\renewcommand{\thefigure}{S\arabic{figure}}

\renewcommand{\theequation}{S\arabic{equation}}

\begin{center}
    \huge{{\bf Supplementary Materials}}
\end{center}

\section{Development of interventional prediction models using observational data}
\label{sec:development}

\subsection{Development data}
\label{subsec:model.dev.data}

A cohort of individuals is assumed to be available for development of a model for predictions under interventions. We let $T$ denote the time to the event of interest, measured relative to the time point from which a prediction would be made, and $C$ denotes the censoring time. For individual $i$ the observed end of follow-up is $T_i^*=\mathrm{min}(T_i,C_i)$, and $D_i$ is the event indicator. We focus on a setting in which each individual in the cohort is observed at regular time points (e.g. study visits) $k=0,1,\ldots$ up to the event or censoring time. Time-dependent covariates $L_k$ and treatment status $A_k$ are recorded at each visit. Additional prognostic variables $P$ are recorded at baseline. Events and censorings are assumed to be observed in continuous time (i.e. not just at the visit times). We let  $X=\{L_0^*,P\}$ denote the baseline characteristics to be used when estimating the risk, where $L_0^*$ denotes a subset of $L_0$, which could be all or none of $L_0$. We focus on the setting where all individuals are untreated before time 0 ($A_{0^-}=0$) (incident users), but they may follow any treatment pattern from time zero onwards, with the treatment pattern followed depending on both baseline and time-dependent covariates $L$. 

The assumed data structure is as illustrated in the DAG in Figure \ref{fig:dag} in the main text. That DAG refers to the assumed structure of the validation data, but we assume it also applies to the development data here. However, we emphasise that there is no requirement for the development and validation datasets to be of the same form. Main Figure \ref{fig:dag} depicts the presence of time-dependent confounding, which arises when there are time-dependent covariates predictive of the outcome that also inform treatment initiation or continuation, and which are affected by past treatment.

\subsection{Development methods}

Below we outline the MSM-IPTW approach and cloning-censoring-weighting approach. Our focus is on two sustained treatment strategies: (i) never initiating treatment, which we refer to as the \emph{never treated} strategy; (ii) initiating treatment at time 0 and sustaining treatment thereafter, which we refer to as the \emph{always treated} strategy.

\emph{Development using MSM-IPTW}
\label{subsec:model.dev.MSM}

The MSM-IPTW approach involves first specifying a marginal structural hazard model, which is a model for the hazard for counterfactual event times given a longitudinal treatment strategy. Because we wish to obtain estimates of risk conditional on baseline covariates $X$, these should also be conditioned on in the MSM. We let $T^{\underline{a}_0}$ denote the counterfactual event time under the treatment strategy $\underline{a}_0$, which denotes treatment status from time 0 onwards. 
The MSM for the hazard can take any form. Cox models \citep{Cox:1972} are often used, and under this model a general form for the MSM for the hazard under the treatment strategy $\underline{a}_0$ is
\begin{equation}
    h_{T^{\underline{a}_0}} (t|X;\bm \beta)=h_0(t)\exp\left\{g(\bar{a}_{\lfloor t\rfloor};\beta_A)+\beta_X^{\top}X\right\}
    \label{eq:msm}
\end{equation}
where $\lfloor t\rfloor$ denotes the time of the most recent visit before time $t$, $\bar{a}_{\lfloor t\rfloor} =\{a_0,a_1,\ldots,a_{\lfloor t\rfloor}\}$ denotes treatment history up to time $\lfloor t\rfloor$, and $g(\bar{a}_{\lfloor t\rfloor};\beta_A)$ denotes a function of that history. The MSM could be extended to incorporate interactions between treatment and components of $X$ that are known or suspected treatment effect modifiers.

When there is time-dependent confounding, as in main Figure \ref{fig:dag}, fitting the MSM directly using the observed development data will result in biased estimates of the parameters. However, the MSM can be estimated using IPTW under the assumptions of conditional sequential exchangeability (no unmeasured confounding), consistency, and positivity. These assumptions are as stated in the main text section on \emph{Artificial censoring and inverse probability weighting}. The weight for a given individual at time $t$ is the inverse of the conditional probability of having had their observed treatment pattern up to time $t$ given their past treatment status and time-dependent covariate history. Under the assumed data structure illustrated, the covariate history required to control confounding is $\bar{L}_t$, and the weights are
   \begin{equation}
  \label{eqn:weights.supp}
     \prod_{s=0}^{\lfloor t \rfloor}\frac{1}{\Pr(A_s=a_s|\bar A_{s-1}=\bar a_{s-1},\bar L_s)}.
 \end{equation}
Stabilized weights can be used, for example of the form
   \begin{equation}
  \label{eqn:stabweights.supp}
     \prod_{s=0}^{\lfloor t \rfloor}\frac{\Pr(A_s=a_s|\bar A_{s-1}=\bar a_{s-1},X)}{\Pr(A_s=a_s|\bar A_{s-1}=\bar a_{s-1},\bar L_s)}.
 \end{equation}
Any baseline variables that are confounders and that are conditioned on in the model in the numerator of the stabilized weights must be included in the MSM. In this case, as the MSM includes $X$, the model in the numerator of the stabilized weights can include all or a subset of the variables in $X$. A detailed discussion of weight stabilisation is given in \cite{HernanRobins:2020} (Section 12.3). The weights are typically estimated using logistic regressions for treatment status at each visit time, or a pooled logistic regression across time points.

The MSM in (\ref{eq:msm}) can then be fitted using the time-updated weights. Some applications of the MSM-IPTW approach have focused on a discrete time setting and use pooled logistic regression models fitted over series of time intervals, which is asymptotically equivalent to a Cox regression when the discrete time-periods get small. Risk under the treatment strategy $\underline a_0$, as defined in main text equation (\ref{eq:risk}), can then be estimated using the relation
\begin{equation}
    R^{\underline a_0} (\tau|X;\bm \beta)=1-\exp\left\{-\int_{0}^{\tau}h_{T^{\underline a_0}} (u\mid X;\bm \beta)du\right\}.
    \label{eq:risk.msm}
\end{equation}

\emph{Development using the cloning-censoring-weighting approach}
\label{subsec:model.dev.clonecensweigh}

The cloning-censoring-weighting approach is a closely related alternative to MSM-IPTW that focuses on estimation of risks under a restricted set of treatment strategies of interest, whereas the MSM in equation (\ref{eq:msm}) enables estimation under any longitudinal treatment strategy. The first step in this approach (`cloning') is to create a copy of the development dataset for each strategy of interest. In the `censoring' step, the dataset corresponding to a given treatment strategy $\underline a_0$ is modified such that each individual’s follow-up is used for the duration for which their treatment status is consistent with the strategy $\underline a_0$, by censoring people when their observed treatment status deviates from that strategy. For example, if we were interested in the \emph{never treated} ($\underline a_0=\mathbf 0$) and \emph{always treated} ($\underline a_0=\mathbf 1$) strategies we would create two copies of the data. In the \emph{never treated} dataset individuals would be censored at the visit time at which they initiate treatment ($A_k=1$), if that occurs, and individuals with $A_0=1$ would have zero follow-up. Similarly, in the \emph{always treated} dataset individuals would be censored at the visit time at which they stop taking treatment ($A_k=0$), meaning that individuals with $A_0=0$ would have zero follow-up. Because time-dependent covariates are associated with changes in treatment status and with the outcome, the `artificial' censoring is informative, and this is addressed in the analysis by using inverse probability of artificial censoring weighting (IPACW). In the dataset for treatment strategy $\underline a_0$, the IPACW at time $t$ for an individual who has not yet had the event or been artificially censored is the inverse of the probability of having sustained treatment strategy $\underline a_0$ up to time $t$, meaning that the weights for individuals not artificially censored are the same as in (\ref{eqn:weights.supp}) or (\ref{eqn:stabweights.supp}). 
To obtain estimates of conditional risks a model for the conditional hazard $h_{T^{\underline a_0}} (t|X)$ could be fitted for each treatment strategy of interest, i.e. separately for each of the modified datasets. Alternatively, combined models could be fitted. For example, if we were interested in the \emph{never treated} and \emph{always treated} strategies the MSM could be a stratified Cox model of the form
\begin{equation}
    h_{T^{\underline a_0=a}} (t|X;\bm \beta)=h_{0a}(t)\exp\left\{\beta_X^{\top}X\right\},\qquad a=0,1.
    \label{eq:msm.cens}
\end{equation}
Risks under the treatment strategies of interest, as defined in main text equation (\ref{eq:risk}), can then be estimated using the relation in (\ref{eq:risk.msm}).


We emphasise that the weights used in the development step for an interventional prediction model are different from those used in the validation step, as treatment assignment may be different in the development and validation datasets. Even if the data follow the same structure, our validation approach uses unstabilised weights.


\section{Counterfactual cumulative dynamic AUCt}
\label{sec:auc.suppl}

Several versions of AUCt have been proposed \citep{heagerty_survival_2005}. We focus on the cumulative/dynamic AUCt, which measures discrimination \emph{at} a particular time point $t$. A corresponding c-index, which summarizes discrimination over a range of follow up times \emph{up to} a time horizon $\tau$, was discussed in the section on \emph{Counterfactual discrimination} in main text.

More specifically, the cumulative/dynamic AUCt assesses concordance of predictions for pairs comprising cumulative cases (subjects with $T\leq t$), and dynamic controls (subjects with $T>t$) \citep{heagerty_survival_2005}. Under treatment strategy $\underline{a}_0$ the cumulative/dynamic AUCt is defined as 
\begin{equation}
    AUC^{\underline{a}_0}_{C/D}(t)=\Pr(\hat R^{\underline{a}_0}(t|X_i)>\hat R^{\underline{a}_0}(t|X_j)|T^{\underline{a}_0}_i\leq t,T^{\underline{a}_0}_j>t).
\end{equation}
We propose the following weighted estimator for $AUC^{\underline{a}_0}_{C/D}(t)$: 
\begin{equation}
    \hat{AUC}_{C/D}^{\underline{a}_0}(t)=\frac{\sum_{i=1}^{n} \sum_{j=1}^{n} I(\hat R^{\underline{a}_0}(t|X_i)>\hat R^{\underline{a}_0}_j(t|X_j))\mathrm{comp}^{(3)}_{\underline{a}_0,ij}(t)\hat{W}^{(3)}_{\underline{a}_0ij}}
{\sum_{i=1}^{n} \sum_{j=1}^{n} \mathrm{comp}^{(3)}_{\underline{a}_0,ij}(t)\hat{W}^{(3)}_{\underline{a}_0,ij}}
\end{equation}
where $\mathrm{comp}^{(3)}_{\underline{a}_0,ij}(t)=I(\tilde{T}_{\underline{a}_0i}\leq t, \tilde{T}_{\underline{a}_0j}>t, \tilde{D}_{\underline{a}_0i}=1)$ indicates whether the pair of subjects $(i,j)$ is comparable at time $t$ in $V^{\underline{a}_0}$, and $\hat{W}^{(3)}_{\underline{a}_0,ij}=\hat{G}_{\underline{a}_0c}^{-1}(\tilde{T}_{\underline{a}_0i}|{L}_{i})\hat{G}_{\underline{a}_0c}^{-1}(t|{L}_{j})$ is the weight of the pair. 

\section{Simulation Plan}
\label{sec:sim.plan.suppl}

We follow the general recommendations of \cite{Morris:2019} on the conduct of simulation studies for evaluating statistical methods. 

\emph{Aim}
\label{subsec:sim.plan.aim}

The aim of this simulation study is to evaluate the performance of our proposed methods for assessing predictions under interventions using longitudinal observational data. Specifically, we aim to investigate whether the proposed counterfactual performance measures (calibration, discrimination, and Brier score) give unbiased estimates of the true performance measures that would be observed if we had a perfect validation set where everyone followed the treatment strategy of interest. Our interest includes the ability of the methods to detect poor predictive performance. 

\emph{Data generating mechanisms}
\label{subsec:sim.plan.dgm}

We generate development datasets, observational validation datasets and `perfect' validation datasets. The perfect validation datasets allow construction of the true performance measures. All datasets are generated according to the longitudinal structure illustrated in the DAG in Figure \ref{fig:dag.suppl}, with treatment $A_k$ and time-dependent covariate $L_k$ that introduces time-dependent confounding being generated at five visits ($k=0,1,2,3,4$), alongside the continuous time to event. Compared to the general DAG introduced in the main text, we do not use $P$ in the simulation, meaning that in this case $X$, the conditioning set in the interventional prediction model, only contains $L_0$. The DAG in Figure \ref{fig:dag.suppl} additionally accounts for an unobserved variable $U$ that affects $L$ and $Y$, to introduce some realistic unexplained randomness in the longitudinal marker and the outcome. Note that U does not affect $A$, so it does not introduce confounding.

Data generation requires models for $A_k|\bar A_{k-1},\bar L_k$, for $L_k|\bar A_{k-1},\bar L_{k-1},U$, and for the hazard for the event at time $t$ conditional on the treatment and covariate history, denoted $h(t|A_{\lfloor t \rfloor},L_{\lfloor t \rfloor},U)$. 

Three main scenarios are considered. In Scenario 1, the development and validation datasets are generated under the same model for the conditional hazard $h(t|A_{\lfloor t \rfloor},L_{\lfloor t \rfloor})$. In Scenario 2 the development dataset has a higher baseline hazard than the validation dataset, but the form of the hazard model is otherwise the same. In Scenario 3 the development and validation datasets are generated under the same mechanism as in Scenario 1, but the predictions in the validation data are obtained using an error prone version of $L_0$, denoted $L_0^*$. Scenarios 2 and 3 mimic settings where we expect poor predictive performance. 

Three additional scenarios were considered in which we assessed our method’s performance under respective violation of the assumptions of positivity (Scenarios 4a-b), conditional exchangeability (Scenarios 5a-b), and correct specification of the weight model (Scenarios 6a-d).

For each scenario we consider data generating mechanisms using an additive hazards model \citep{Aalen:1989} for $h(t|A_{\lfloor t \rfloor},L_{\lfloor t \rfloor},U)$ and using a proportional hazards model for $h(t|A_{\lfloor t \rfloor},L_{\lfloor t \rfloor},U)$. The data generating mechanisms are summarised in detail in Table \ref{tab:sim.datagen} for the additive hazards model and in Table \ref{tab:sim.datagen.cox} for the proportional hazards model. The reason for considering a data generating mechanism using an additive hazards model is that it enables us to fit a correctly specified MSM during model development for $h_{T^{\underline{a}_0}} (t|L_0)$, as the form of $h_{T^{\underline{a}_0}} (t|L_0)$ is then also an additive hazards model \citep{Keogh:2021}. This enables us to assess the proposed validation methods in an `ideal' scenario in which the development model is correctly specified and the development and validation data are generated under the same mechanism (Scenario 1).
When $h(t|A_{\lfloor t \rfloor},L_{\lfloor t \rfloor})$ is a conditional proportional hazards model, the MSM for $h_{T^{\underline{a}_0}} (t|L_0)$ is no longer a proportional hazards model and in fact is of a complex non-standard form. However, a data generating mechanism based on a proportional hazards model is more widely familiar. Including a data generating scenario using a proportional hazards model also provides one example of a situation in which the development model will be mis-specified so we would expect the risks estimated from this model to be biased estimates of the true risks under interventions in the validation data. This model mis-specification should be reflected in the performance measures.

We chose a sample size of n=3000 in each simulation run. This choice was motivated by having at least 80\% power to accurately (within 5\% margin of error) estimate the overall outcome proportion using the artificially censored data in each scenario \citep{riley_calculating_2020}. We performed 1000 simulation runs.

\emph{Estimands}
\label{subsec:sim.plan.estimands}

The estimands of interest are the counterfactual measures of predictive performance for assessment of predictions under interventions introduced in the main paper (equation (\ref{eq:risk}) with $X=L_0$). 
We obtain estimates of risk for each individual in the observational validation dataset under the treatment strategies of interest, in our case the \emph{always treated} and \emph{never treated} strategies, at time horizons $\tau=1,2,3,4,5$. Our aim is not to consider the accuracy or precision of these risk estimates, but to assess their predictive performance. To obtain `true' values of predictive performance, we extend each generated observational validation dataset into two `perfect' validation datasets, one for the \emph{always treated} and one for the \emph{never treated} strategy. These perfect validation datasets inherit the baseline values $L_0$ and $U$ from the observational validation dataset. Later values $L_k, k=1,2,3,4$ and the times to event are generated assuming all patients follow the \emph{always treated} or the \emph{never treated} strategy. These perfect validation datasets constitute the ideal validation setting for the predictions under interventions: as all patients follow the strategy of interest, standard estimators for performance measures 
can be used. 
In the observational validation dataset we calculate the performance measures listed in Table \ref{tab:estimands} and we compare these to the measures obtained from the perfect validation datasets.

\emph{Methods}
\label{subsec:sim.plan.methods}

An interventional prediction model is fitted in the development data using an MSM-IPTW analysis as outlined in Section \ref{subsec:model.dev.MSM}. The development model is used to obtain estimates of risks for each person at time horizons $\tau=1,2,3,4,5$ in the observational validation data under the \emph{always treated} strategy and under the \emph{never treated} strategy. In main Scenarios 1 and 2 (and additional Scenarios 4-6) this is done using $L_0$ in the validation data. In Scenario 3 the predictions are obtained using an error prone version of baseline measurements $L_0^*$. 

The counterfactual performance measures as described Table \ref{tab:estimands} are estimated from the observational validation data following two approaches. First, we apply the  subset approach as described in the \emph{Introduction} section in the main paper
. This means we apply standard predictive performance measures to the subset of patients in the observational validation data who followed the treatment strategy of interest (\emph{always treated} or \emph{never treated}). The subset is specific to the time horizon over which a certain performance measure is calculated. Second, we apply the proposed artificial censoring and inverse probability weighting approach to assess counterfactual performance. 
Note that the weights used during the validation are estimated in the validation data. As depicted in Tables \ref{tab:sim.datagen}(a) and \ref{tab:sim.datagen.cox}(a), the models used to generate treatment assignment $A_k$ only depend on the latest value of $\bar{L}_k$, $L_k$. Accordingly, the weight models, depicted in Tables \ref{tab:sim.datagen}(c) and \ref{tab:sim.datagen.cox}(c) only condition on this latest value $L_k$ (accept in Scenario 6 where we introduce misspecification in the weight models).

\emph{Performance measures}

For each simulation run, we will calculate the counterfactual performance measures estimated from the observational validation data and the true performance measure calculated from the perfect validation datasets. We depict the estimates graphically and tabulate the mean of their differences (i.e. the bias) and the Monte Carlo standard error of the bias.  

\section{Simulation Results}
\label{sec:sim.results.suppl}

\subsection{Simulation Descriptives}
\label{subsec:sim.results.descr}

\begin{figure}
	\centering
		\includegraphics[scale=0.75]{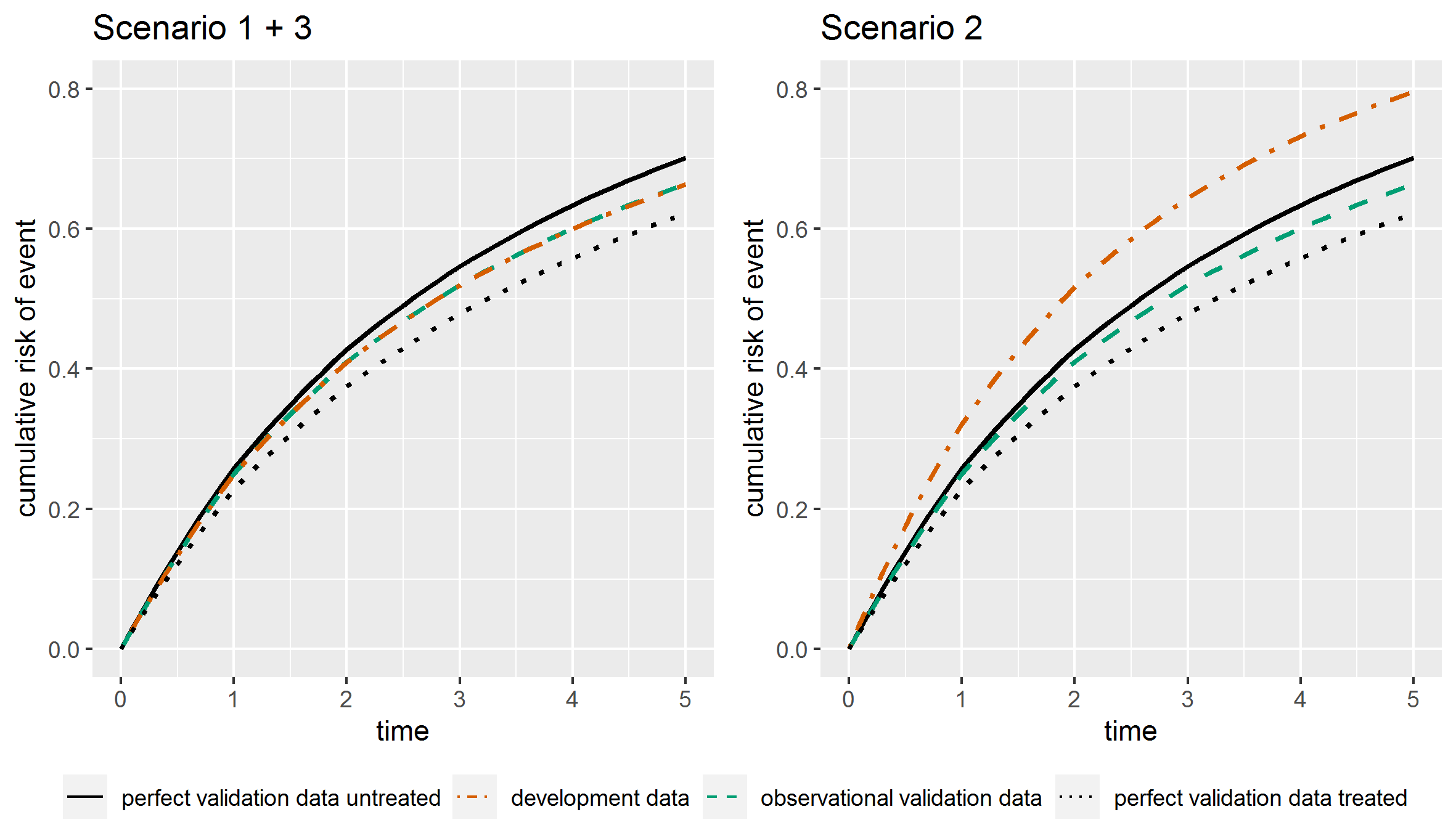}
	    \caption{Marginal risk distribution estimated from the development datasets, observational validation datasets and perfect counterfactual \emph{never treated} and \emph{always treated} datasets generated using the additive hazards model, main Scenarios 1-3. The curves are constructed by averaging over the (one minus) survival curves estimated by the Kaplan-Meier estimator in each of the 1000 simulation runs.}\label{fig:sim_addhaz_KM}
\end{figure}

The marginal risk distributions averaged over the 1000 simulation runs and from the development data, observational validation data and the two perfect validation datasets for the main Scenarios 1-3 generated under the additive hazards model are shown in Figure \ref{fig:sim_addhaz_KM}. In the \emph{never treated} perfect validation data the average risk by time point 5 is 70\% (2100 events). The corresponding marginal risk from the \emph{always treated} perfect validation data is 62\% (1866 events on average by time point 5), meaning that treating all subjects would lower the overall risk by about 8 percentage points compared to not treating any patient. In the development data (for Scenarios 1 and 3) and observational validation data (for Scenarios 1-3), on average 53\% of patients started treatment at some point during follow up. The mix of treated and untreated individuals in the development (Scenarios 1 and 3) and observational validation data led to an overall risk of 66\% by time point 5 (on average 1980 events). In the artificially censored validation data for the \emph{never treated} strategy ($V^0$) on average 1122 events remained in the analysis and in the artificially censored validation data for the \emph{always treated} strategy ($V^1$) 534 events remained. 

\begin{figure}
	\centering
		\includegraphics[scale=0.75]{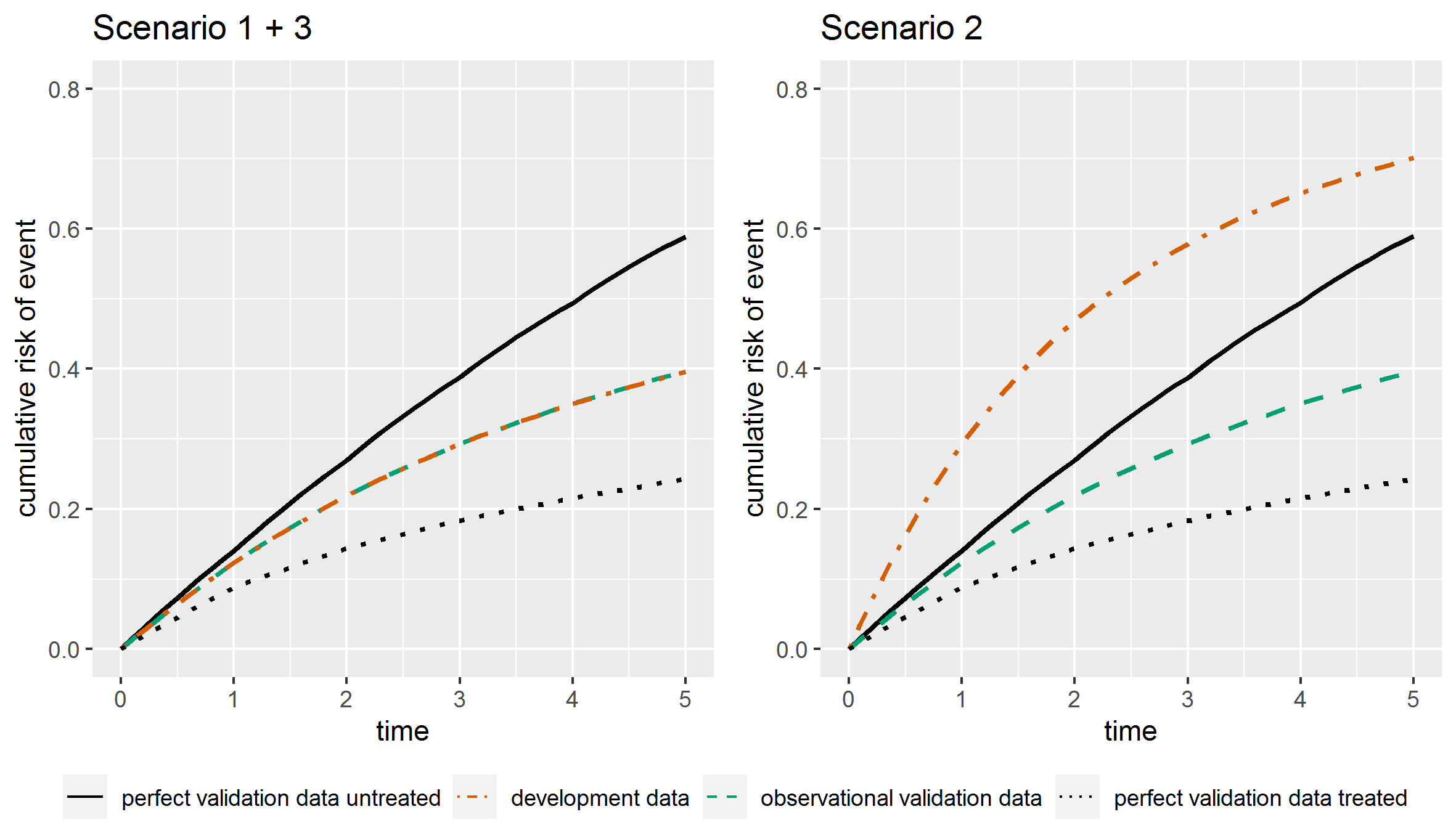}
	    \caption{Marginal risk distribution estimated from the development datasets, observational validation datasets and perfect counterfactual \emph{never treated} and \emph{always treated} datasets generated using the Cox model, main Scenarios 1-3. The curves are constructed by averaging over the (one minus) survival curves estimated by the Kaplan-Meier estimator in each of the 1000 simulation runs.}
     \label{fig:sim_cox_KM}
\end{figure}

The risk distributions for the proportional hazards based scenarios where we assumed a stronger treatment effect, a stronger effect of $L$ and a higher percentage of patients who received treatment, are presented in Figure \ref{fig:sim_cox_KM}. Under the proportional hazards model in the \emph{never treated} perfect validation data the average risk by time point 5 is 59\% (1769 events). In the corresponding \emph{always treated} perfect validation data, the average risk by time point 5 was 24\% (729 events), so 35 percentage points lower. In the development data (Scenarios 1 and 3) and observational validation data, on average 68\% of patients started treatment at some point during follow up. 
The mix of treated and untreated individuals in the development data (Scenarios 1 and 3) and observational validation data led to an average risk of 40\% by time point 5. In the artificially censored validation data for the \emph{never treated} strategy ($V^0$) on average 680 events remained in the analysis and in the artificially censored validation dataset for the \emph{always treated} strategy ($V^1$) 227.

\subsection{Simulation results proportional hazards scenarios}
\label{subsec:sim.results.cox}

\begin{figure}
	\centering
		\includegraphics[scale=0.98]{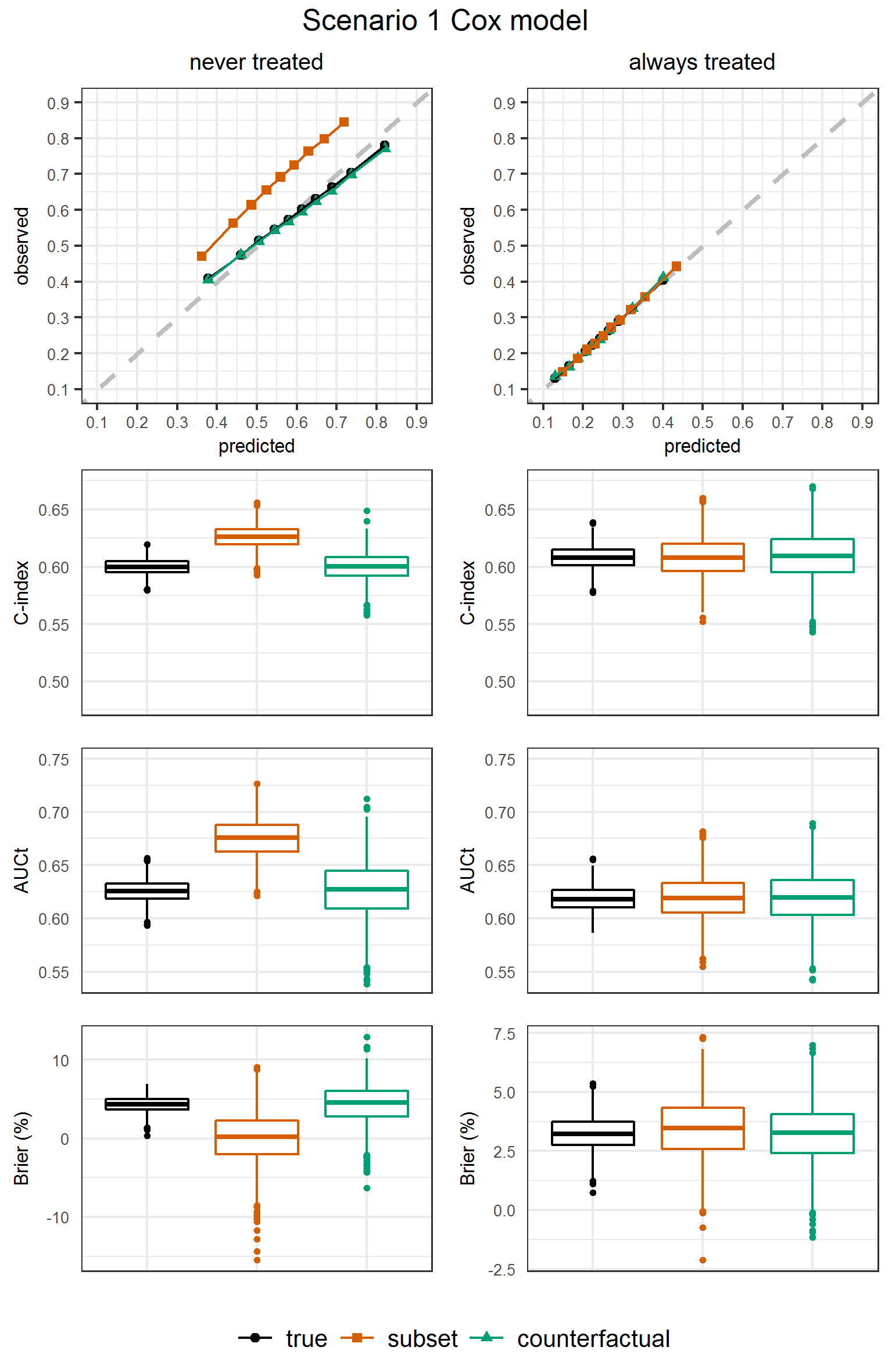}
	    \caption{Simulation results: Cox model Scenario 1. Left panel:  for the \emph{never treated} strategy. Right panel: for the \emph{always treated} strategy. Performance measures were obtained from the perfect validation data (black dots) and estimated from the observational validation data using the subset approach (orange squares) and using the proposed artificial censoring + inverse probability weighted estimators of counterfactual performance (green triangles). Top row: calibration plot showing observed outcome proportions against mean estimated risks by time 5 within tenths of the estimated risks. Second row: c-index truncated at time 5. Third row: cumulative/dynamic area under the ROC curve at time 5. Bottom row: scaled Brier score at time 5. }\label{fig:sim.cox.scenario1.main}
\end{figure}

\begin{figure}
	\centering
		\includegraphics[scale=1.0]{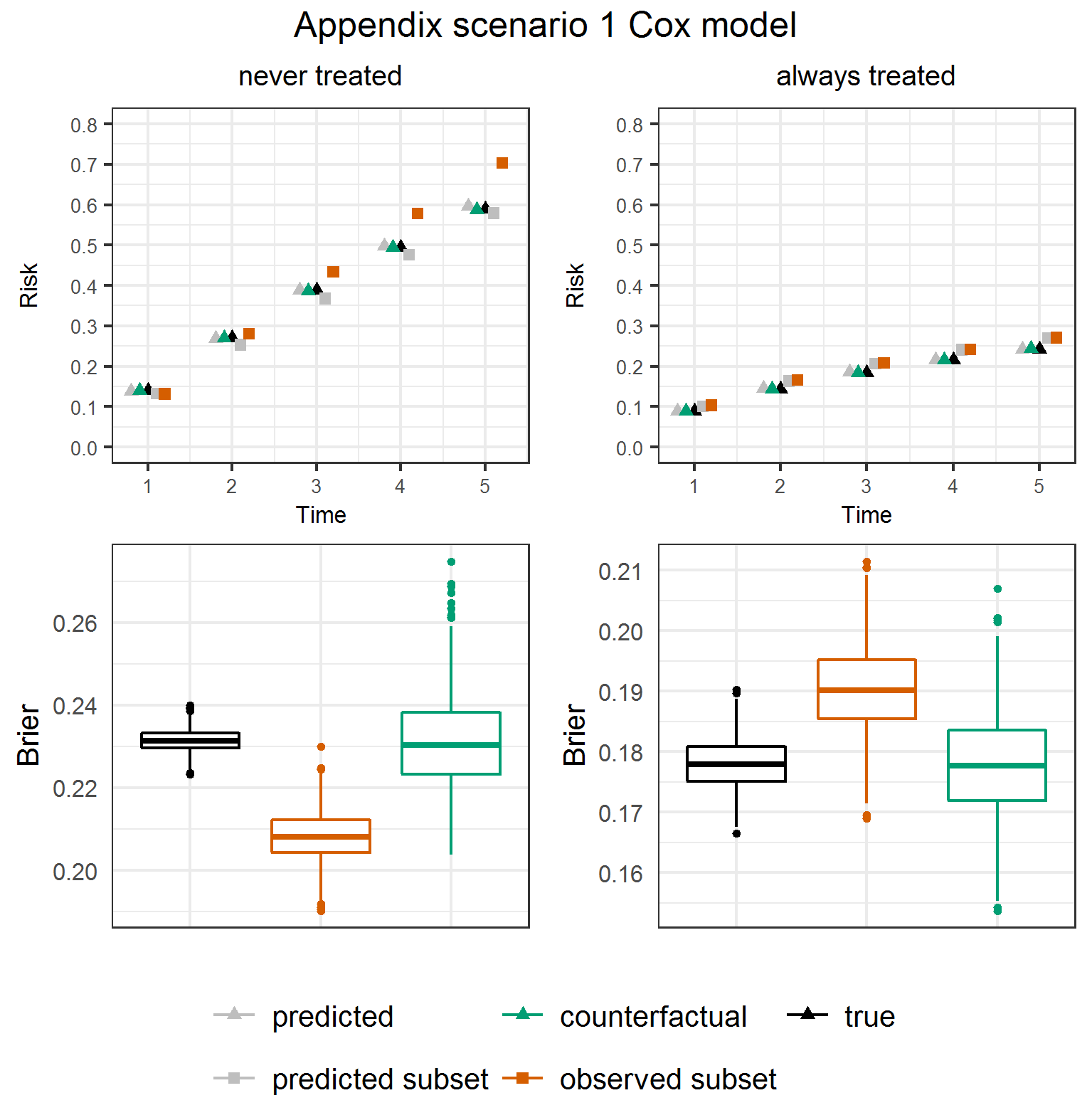}
	    \caption{Simulation results: Cox model Scenario 1. Left panel:  for the \emph{never treated} strategy. Right panel: for the \emph{always treated} strategy. Top row: outcome proportions over time estimated by the prediction model (grey triangles), observed in the perfect validaiton data (black triangles) and estimated from the observational validation data using the proposed artificial censoring + inverse probability weighting estimators for counterfactual performance assessment (green triangles). Estimated and observed outcome proportions using the subset approach are depicted with grey and orange squares.  Bottom row: unscaled Brier score at time 5.}\label{fig:sim.cox.scenario1.appendix}
\end{figure}

\begin{figure}
	\centering
		\includegraphics[scale=0.98]{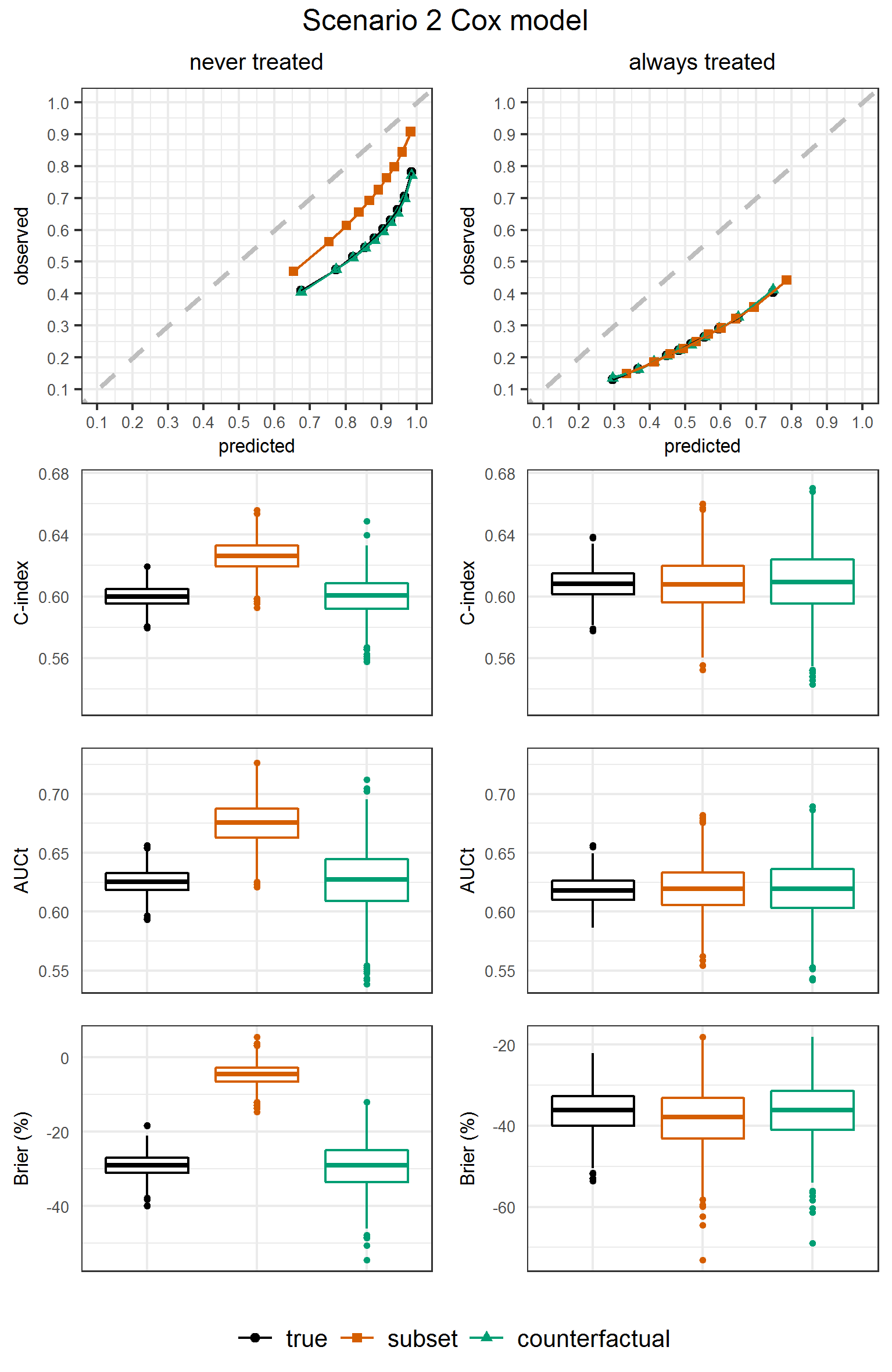}
	    \caption{Simulation results: Cox model Scenario 2. Left panel:  for the \emph{never treated} strategy. Right panel: for the \emph{always treated} strategy. Performance measures were obtained from the perfect validation data (black dots) and estimated from the observational validation data using the subset approach (orange squares) and using the proposed artificial censoring + inverse probability weighted estimators of counterfactual performance (green triangles). Top row: calibration plot showing observed outcome proportions against mean estimated risks by time 5 within tenths of the estimated risks. Second row: c-index truncated at time 5. Third row: cumulative/dynamic area under the ROC curve at time 5. Bottom row: scaled Brier score at time 5. }\label{fig:sim.cox.scenario2.main}
\end{figure}

\begin{figure}
	\centering
		\includegraphics[scale=1.0]{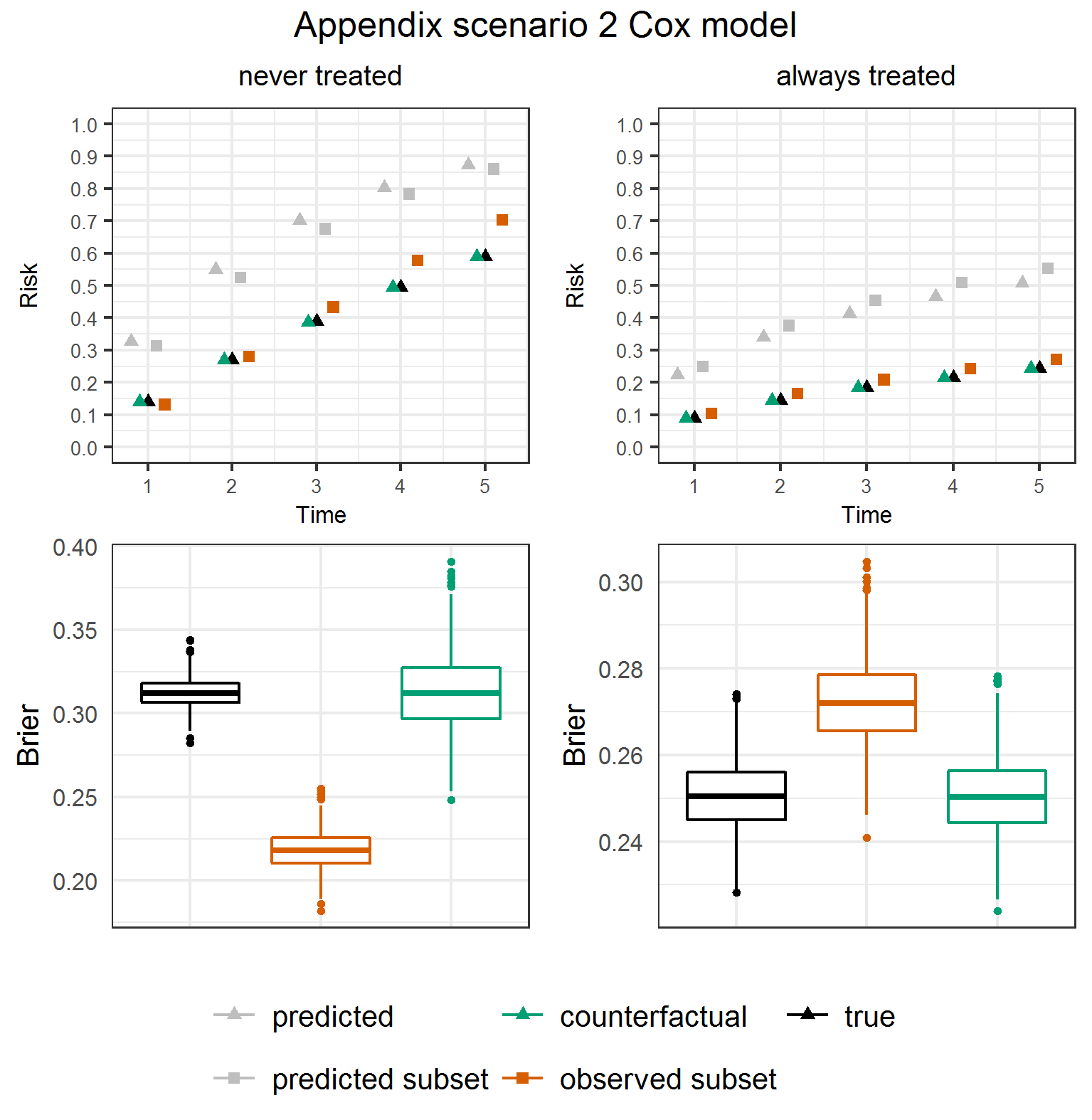}
	    \caption{Simulation results: Cox model Scenario 2. Left panel:  for the \emph{never treated} strategy. Right panel: for the \emph{always treated} strategy. Top row: outcome proportions over time estimated by the prediction model (grey triangles), observed in the perfect validaiton data (black triangles) and estimated from the observational validation data using the proposed artificial censoring + inverse probability weighting estimators for counterfactual performance assessment (green triangles). Estimated and observed outcome proportions using the subset approach are depicted with grey and orange squares.  Bottom row: unscaled Brier score at time 5.}\label{fig:sim.cox.scenario2.appendix}
\end{figure}

\begin{figure}
	\centering
		\includegraphics[scale=0.98]{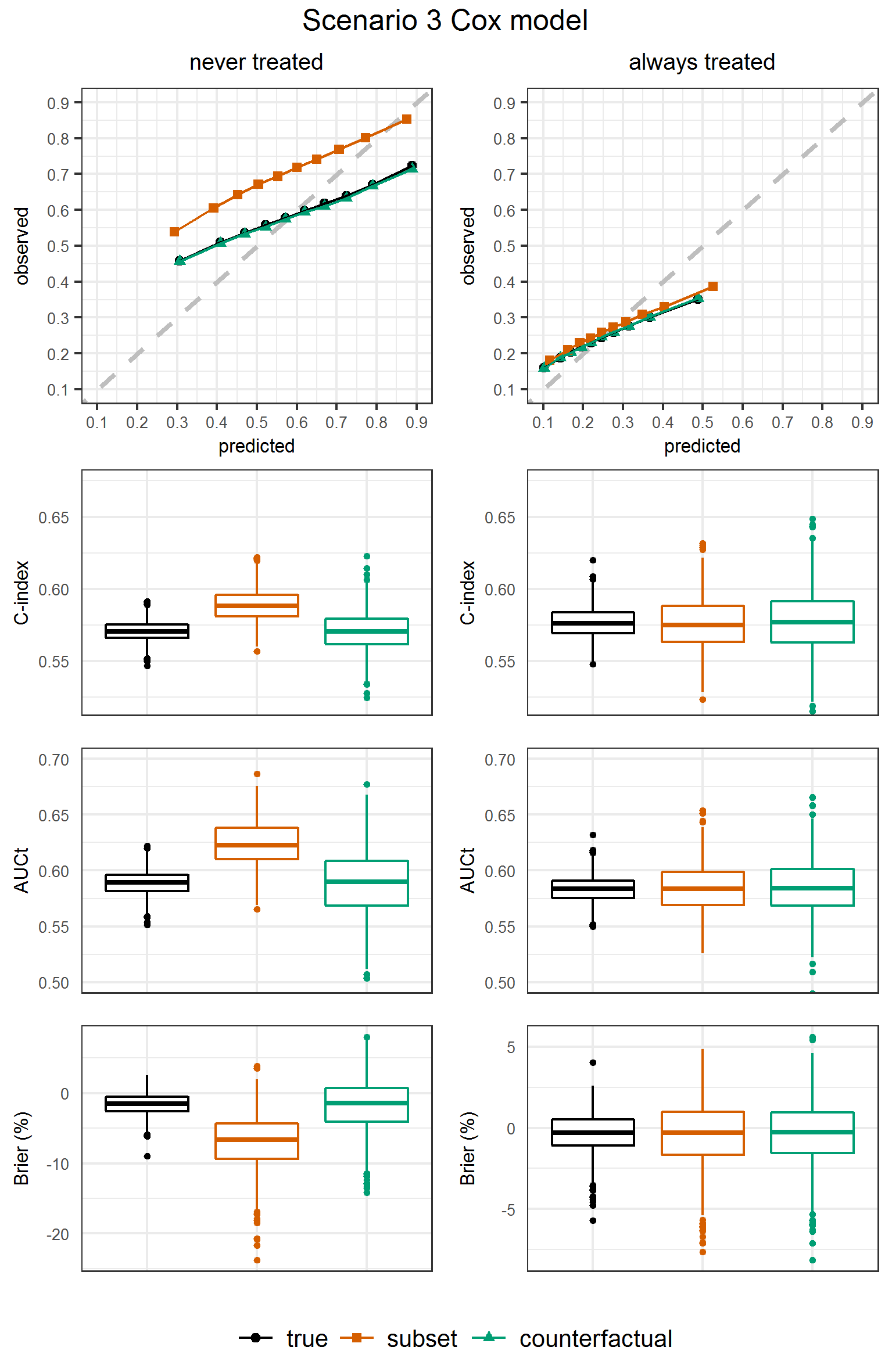}
	    \caption{Simulation results: Cox model Scenario 3. Left panel:  for the \emph{never treated} strategy. Right panel: for the \emph{always treated} strategy. Performance measures were obtained from the perfect validation data (black dots) and estimated from the observational validation data using the subset approach (orange squares) and using the proposed artificial censoring + inverse probability weighted estimators of counterfactual performance (green triangles). Top row: calibration plot showing observed outcome proportions against mean estimated risks by time 5 within tenths of the estimated risks. Second row: c-index truncated at time 5. Third row: cumulative/dynamic area under the ROC curve at time 5. Bottom row: scaled Brier score at time 5. }\label{fig:sim.cox.scenario3.main}
\end{figure}

\begin{figure}
	\centering
		\includegraphics[scale=1.0]{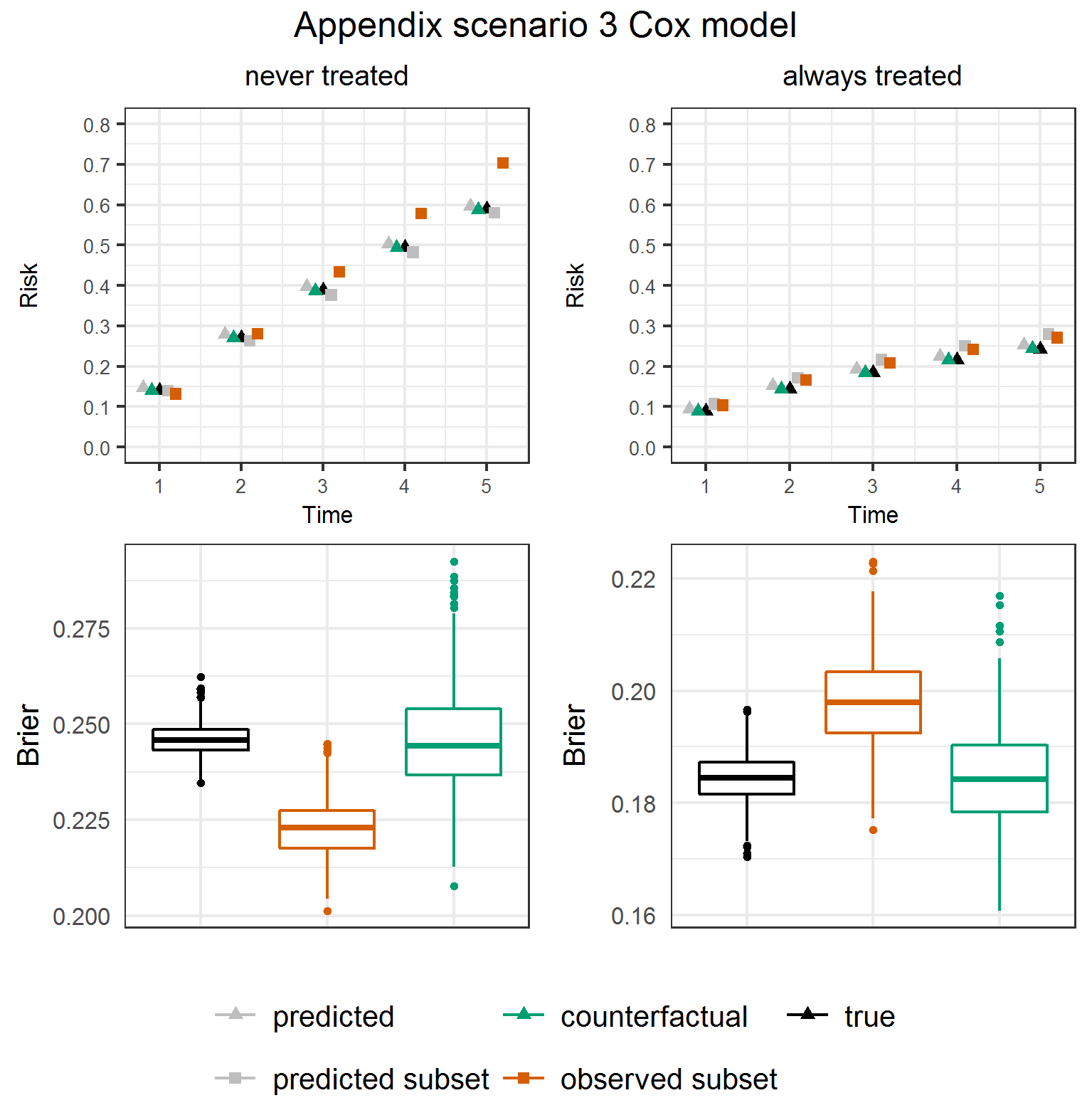}
	    \caption{Simulation results: Cox model Scenario 3. Left panel:  for the \emph{never treated} strategy. Right panel: for the \emph{always treated} strategy. Top row: outcome proportions over time estimated by the prediction model (grey triangles), observed in the perfect validaiton data (black triangles) and estimated from the observational validation data using the proposed artificial censoring + inverse probability weighting estimators for counterfactual performance assessment (green triangles). Estimated and observed outcome proportions using the subset approach are depicted with grey and orange squares.  Bottom row: unscaled Brier score at time 5.}\label{fig:sim.cox.scenario3.appendix}
\end{figure}

Results for the scenarios in which data were generated and analysed using a proportional hazards model are presented in Tables \ref{tab:cox.sc1}, \ref{tab:cox.sc2}, \ref{tab:cox.sc3} and \ref{tab:cox.sc456} and Figures \ref{fig:sim.cox.scenario1.main} to \ref{fig:sim.cox.scenario3.appendix}. 
As explained in Section \ref{sec:development}, using the proportional hazards model during model development will lead to a mis-specified MSM based on a Cox proportional hazards model and this is reflected in the true calibration curve not lying perfectly on the diagonal in Scenario 1. Results for the proportional hazards model confirm the unbiasedness of the proposed estimators when the necessary assumptions are met (Scenarios 1-3) and show somewhat stronger biases (compared to the additive hazards based data generation) for the scenarios with deliberately introduced violations of causal assumptions (Scenarios 4-6). This is attributable to the stronger treatment effect and stronger effects of $L$ in these proportional hazards based scenarios. The bias of the proposed estimators was smaller than that of the naive subset method in 52/64 (81\%) of times in the proportional hazards based scenarios (Table \ref{tab:cox.sc456}). 


\subsection{Discussion of the size and direction of bias when using the subset method}
\label{subsec:sim.results.subset}

\begin{figure}
	\centering
		\includegraphics[scale=1.2]{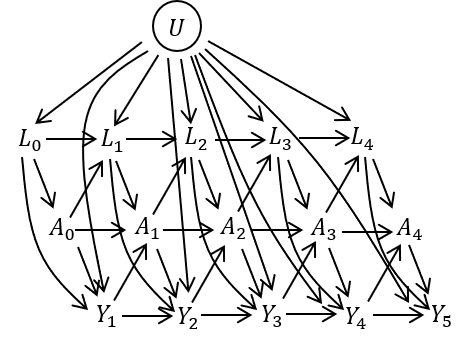}
			\caption{Directed acyclic graph (DAG) illustrating the  data generating mechanism for the simulation study, for treatment $A$, time-dependent covariates $L$, and discrete time outcome $Y$. The DAG is illustrated for a discrete-time setting where $Y_k=I(k-1 \leq T< k)$ is an indicator of whether the event occurs between visits $k-1$ and $k$. If the DAG is extended by adding a series of small time intervals between each visit, at which events are observed, then we approach the continuous time setting. The covariates $L_k$ are time-dependent confounders as they inform treatment initiation or continuation, are predictive of the outcome, and are also affected by past treatment.}\label{fig:dag.suppl}
\end{figure}

Bias in the estimates using the subset method were more pronounced and sometimes in opposite directions for the \emph{never treated} strategy compared to the \emph{always treated} strategy. Here we explain the mechanisms behind this. Under the \emph{never treated} strategy, the subset is obtained by excluding subjects if they start treatment at any time point during the follow-up time of interest, meaning that we have `selection based on the future'. Notably, subjects who experience the event can no longer be excluded after that, introducing a type of `immortal time bias'. This leads to an over representation of events in the resulting subset under the \emph{never treated} strategy and explains the overestimation of observed outcome proportions by the subset method under that strategy. For the \emph{always treated} strategy, selection into the subset is based only on the first visit. Patients are excluded if they do not start treatment directly, but remain in the analysis after that because under our data generating mechanism individuals always continue treatment after it is initiated. The exclusion decision is made at time zero before any events are recorded, meaning that there is no `immortal time' in the \emph{always treated} subset. Selection into the subset in the \emph{always treated} strategy is however not completely at random. Due to the positive relation between $L_0$ and $A_0$, the subjects who receive treatment at time 0 and who thus stay in the subset under the \emph{always treated} strategy on average have higher and more homogeneous underlying risk compared to the counterfactual \emph{always treated} data (Figure \ref{fig:dag.suppl}). With a more homogeneous risk distribution in the subset under the \emph{always treated} strategy compared to the perfect validation data under the \emph{always treated} strategy, one can expect measures of discrimination to decrease \citep{diamond_what_1992, gail_criteria_2005, pajouheshnia_accounting_2017}. This explains the underestimation of discrimination indices seen with the subset approach for this strategy. Discrimination indices for the \emph{never treated} strategy are overestimated by the subset method. This would be consistent with a more heterogeneous underlying risk distribution in the subset used under the \emph{never treated} strategy compared to the risk distribution in the \emph{never treated} perfect validation data. Low risk subjects, who corresponds to low $L$ values, are at low risk of treatment initiation and are thus more often retained in the subset for the \emph{never treated} strategy. High risk subjects, corresponding to high $L$ values, may be excluded from this subset if they start treatment but we also noted that if they experience an event they are retained in the subset for the \emph{never treated} strategy (Figure \ref{fig:dag.suppl}). Apparently, these two processes in effect lead to a more heterogeneous risk distribution explaining the overestimation of discrimination indices by the subset method under the \emph{never treated} strategy.

\section{Liver transplant application}
\label{sec:application.suppl}

\subsection{Data overview}
\label{subsec:application.suppl.data}

This study used data from the Scientific Registry of Transplant Recipients (SRTR). The SRTR data system includes data on all donors, wait-listed candidates, and transplant recipients in the US, submitted by the members of the Organ Procurement and Transplantation Network (OPTN). The Health Resources and Services Administration (HRSA), U.S. Department of Health and Human Services provides oversight to the activities of the OPTN and SRTR contractors. The data reported here have been supplied by the Hennepin Healthcare Research Institute (HHRI) as the contractor for the Scientific Registry of Transplant Recipients (SRTR). The interpretation and reporting of these data are the responsibility of the author(s) and in no way should be seen as an official policy of or interpretation by the SRTR or the U.S. Government.

For this study we restricted to individuals who joined the liver transplant waitlist between 1 January 2014 and 30 April 2019, due to a change in the organ allocation policy in 2019. Administrative censoring was applied at the earlier of 3 years after joining the waitlist or at 30 April 2019. We excluded people with missing information on time-fixed variables (mostly underlying disease group was missing). We also excluded individuals with missing data for time dependent variables, as the number was very small. The data include date of receiving a transplant, date of death (pre- or post-transplant), and date of and reason for removal from the waitlist. The reason for being removed from the waitlist can be due to worsening health status, due to improvement in health status or for ``other'' reasons. We consider a composite outcome of death or removal from the transplant waitlist due to worsening health status. For convenience below we refer to removal due to improvement in health status or for ``other'' reasons collectively as improvement-based-removal. 

Before exclusions due to missing data, the data included 50552 individuals. We excluded 7108 (14.1\%) individuals with missing information on their disease group. A further 242 individuals were excluded because they had missing data on baseline covariates of region, diabetes status or BMI. Lastly we excluded 12 people who had any missing values in time-dependent variables at any measurement time. Missingness occurred only for INR, ascites, encephalopathy, and Child Pugh Score grade. After exclusions, the data included 43190 individuals. 

\subsection{Data set-up}
\label{subsec:application.suppl.setup}

We let $s$ denote time in days since joining the waitlist. The two interventions under which we aim to make predictions are: (1) receiving a liver transplant at time $s$; (2) not receiving a transplant at time $s$ or in the future. The two interventions could be applied at any time $s$ that an individual is on the waitlist pre-transplant. We therefore wish to be able to make predictions from any time $s$ from the time of joining the waitlist and we consider $0\leq s<1096$ (where 1096 days is approximately 3 years).

To enable development of predictions under the two interventions at any time $s$ after joining the waitlist we create two new datasets. We let $D_1$ denote a dataset formed of individuals who receive a transplant within 3 years of joining the waitlist, followed-up from the time of transplant onwards. We also created datasets starting at a series of landmark times from the time of joining the waitlist. We used landmark times at 90-day intervals up to 3 years, giving 13 landmark datasets starting at times $s=0,90,180,\ldots,900,990,1080$. The dataset starting at landmark time $s$ includes individuals who remain on the waitlist at time $s$ - that is, people who have not had a transplant up to time $s$, who have not had the composite event up to time $s$, who have not been removed from the waitlist due to improvement or for ``other'' reasons up to time $s$ (improvement-based-removal), and who have not been administratively censored up to time $s$. The landmark datasets are combined into a single stacked dataset, denoted $D_0$. Individuals can contribute to more than one landmark dataset in $D_0$, and individuals who have a transplant within 3 years of joining the waitlist contribute to both $D_1$ and $D_0$. In dataset $D_1$ time zero is the day of transplant. In dataset $D_0$ time zero is the landmark time. We let $D$ denote datasets $D_1$ and $D_0$ combined. When an individual appears more than once in $D$ we treat the different contributions as separate observations for the analysis, referred to as `person-landmark observations'. To facilitate this we generate a new ID-number ``\texttt{id.$s$}'', which refers to a person-landmark observation, where \texttt{id} denotes the original anonymised unique identifier, and $s$ denotes the landmark time (for $D_0$) or the time of transplant (for $D_1$).

The combined dataset $D$ was divided randomly into a 70\% sample used for model development and a 30\% sample used for the validation, with the sampling being performed on the basis of person-landmark observations using the unique identifiers ``\texttt{id.$s$}''. We let $D^{dev}=\{D_0^{dev},D_1^{dev}\}$ denote the development data and $D^{val}=\{D_0^{val},D_1^{val}\}$ denote the validation data.

Dataset $D_1^{dev}$ includes 16605 individuals who had a transplant within 3 years of joining the waitlist. Among these individuals there were 1748 composite events within 3 years of post-transplant follow-up. The remaining individuals were censored, either because they were still alive after 3 years of follow-up or because of end of follow-up at 30 April 2019. Dataset $D_0^{dev}$ includes 100192 person-landmark observations, which includes some individuals counted in multiple landmark datasets (there are 37210 unique individuals in $D_0^{dev}$). Out of the 100192 person-landmark observations, there were 16717 composite events (death without a transplant or removals from the waitlist due to worsening health status) within 3 years of the landmark time, 40857 had a transplant within 3 years of the landmark, 18137 had improvement-based-removal, and the remaining 24481 person-landmark observations were censored (pre-transplant) because they were still alive after 3 years of follow-up or because of end of follow-up at 30 April 2019.

Dataset $D_1^{val}$ includes 7146 individuals who had a transplant within 3 years of joining the waitlist. Among these individuals there were 777 composite events within 3 years of post-transplant follow-up. The remaining individuals were censored, either because they were still alive after 3 years of follow-up or because of end of follow-up at 30 April 2019. Dataset $D_0^{val}$ includes 42820 person-landmark observations, which includes some people counted in multiple landmark datasets (there are 24228 unique individuals in $D_0^{val}$). Out of the 42820 person-landmark observations, there were 7221 composite events (death without a transplant or removals from the waitlist due to worsening health status) within 3 years of the landmark time, 17293 had a transplant within 3 years of the landmark, 7883 had improvement-based-removal, and the remaining 10423 person-landmark observations were censored (pre-transplant) because they were still alive after 3 years of follow-up or because of end of follow-up at 30 April 2019.

The analysis makes use of time-fixed covariates $Z$ and time-dependent covariates, with $L_s$ denoting the time dependent covariates as measured at time $s$ after joining the waitlist. The following time-fixed covariates were included in $Z$: sex, ethnicity (Asian, Black, White, Other), blood group (A, B, AB, O), region of residence (11 regions), disease group (Alcoholic cirrhosis, Hepititis B virus (HBV) cirrhosis, Hepititis C virus (HCV) cirrhosis, Cryptogenic, Hepatocellular carcinoma (HCC), Non-alcohol related steatohepatitis (NASH), Primary biliary cirrhosis (PBC), Primary sclerosing cholangitis (PSC)), diabetes, BMI. The following time-dependent variables were included in $L_s$: age, chronic kidney disease, dialysis (if the patient had dialysis within the week prior to the serum creatinine test), components of the MELD-NA score (creatinine, bilirubin, INR, sodium), albumin, ascites (Absent, Slight, Moderate), encephalopathy (None, 1-2, 3-4), Child Pugh Score grade (A, B, C), number of tumours (0, 1, 2 or more), whether the person has exception points (yes, no), whether the person has exception points due to HCC (yes, no). Further details about how these are used in the analysis are provided below. 

Characteristics of person-landmark observations in the development and validation datasets at time of joining the waiting list ($s=0$) are summarised in Tables \ref{tab:dev.desc} and \ref{tab:val.desc}, and summaries of numbers of events and censorings in the validation datasets are given in Table \ref{tab:val.events}.



\subsection{Development of interventional prediction models}
\label{subsec:application.suppl.dev}

The data set-up described above and our corresponding analysis are related to approaches taken in earlier studies that have investigated the benefits of organ transplant, though earlier studies have focused on estimating average (or conditional average) effects of transplant rather than on predictions under interventions. Our approach is closely related to that of \cite{gong_estimating_2017}, who estimated the effect of liver transplant in the transplanted, stratified by MELD score. Their data set-up differs from ours in that their landmark times were defined in terms of calendar time rather than in terms of time since joining the waitlist. Strohmeier et al. \cite{strohmaier_survival_2022} used a related approach, but they set their landmark times at each transplant time (relative to moment of joining the waitlist), and their approach is in turn related to the sequential stratification approach of \cite{schaubel_sequential_2006} and \cite{schaubel_estimating_2009}. In this application, our strategies of interest for prediction under interventions do not require assumptions about resource constraint that is faced in organ transplantation. However, in future work other treatment strategies of interest for prediction under interventions may require consideration of the resource constraint, for example the strategy of `waiting until the next donor organ becomes available'. 

\emph{Prediction under transplant}

We developed separate models for prediction under the two interventions. Dataset $D^{dev}_1$ was used to develop a prediction model under the intervention of receiving a liver transplant at time $s$ ($0\leq s<1096$). A Cox model was fitted, with time of transplant as time zero. The model includes as predictors time-fixed variables $Z$ and time-dependent variables $L_s$ as measured just prior to transplant, i.e. $X=\{Z,L_s\}$. All continuous covariates were modelled using restricted cubic splines with 3 knots. The model also included $s$ (time of transplant) as a covariate, which was modelled using a restricted cubic splines with four knots placed at $s$ corresponding to $180, 360, 720, 900$ days. Time-dependent variables were not recorded daily, and we used the last observation carried forward for the values of $L_s$.

\emph{Prediction under no transplant}

Dataset $D^{dev}_0$ was used to develop a prediction model under the intervention of not receiving a liver transplant at time $s$ or in the future ($0\leq s<1096$). All individuals in $D^{dev}_0$ have not received a transplant before their landmark time, but many receive a transplant after the landmark time. We use the censoring-and-weighting approach to obtain predictions under the intervention of interest - that is, not   not receiving a transplant now or in the future. In the censoring step, individuals in $D^{dev}_0$ were censored at the time of receiving a transplant. The resulting modified dataset is denoted $D^{dev,cens}_0$. The artificial censoring at transplant is dependent on time-dependent characteristics of the individual. This is addressed in the analysis using time-dependent inverse probability of artificial censoring weights (IPACW). In addition to the artificial censoring there is administrative censoring due to the end of follow-up at 3 years or at 30 April 2019, and censoring because of improvement-based-removal. The administrative censoring is assumed to be uninformative. However, improvement-based-removal is likely to depend on time-updated individual characteristics that are also associated with the persons subsequent hazard for mortality. We use a second set of weights to address this, referred to as inverse probability of removal weights (IPRW). Once an individual is removed from the waitlist they are assumed not to return. 

For estimating the IPACW and IPRW we divided each individual’s follow-up into periods of length 30 days, starting from the landmark time, enabling use of pooled logistic regression to estimate the weights. Time-dependent covariates are updated at the start of each 30 day period. We let $A_{s+30k}$ denote the transplant status at the start of the $k$th 30-day period following landmark time $s$, with $A_s=0$ denoting transplant status at the landmark time, $L_{s+30k}$ denote the values of time-dependent covariates at the start of period $k$ after landmark time $s$, $L_{s}$ denote the values of time-dependent covariates at the landmark time, and $Z$ denote time-fixed covariates. We let $Q_{s+30k}=1$ denote that an individual is removed from the waitlist at the end of period $k$ after the landmark, and $Q_{s+30k}=0$ otherwise. Time-dependent variables were not recorded daily, and we used the last observation carried forward for the values of $L_s$.

For landmark time $s$, the stabilised IPACW in period $k$ after the landmark is 
\begin{equation}
\mathrm{IPACW}(s,k)=\frac{
\prod_{j=1}^{k}\Pr(A_{s+30j}=0|\bar A_{s+30j-30}=0,Q_{s+30j-30}=0,L_s,Z)}
{\prod_{j=1}^{k}\Pr(A_{s+30j}=0|\bar A_{s+30j-30}=0,Q_{s+30j-30}=0,L_{s+30j},Z)},
\end{equation}
and the stabilised IPRW in period $k$  after the landmark is 
\begin{equation}
\mathrm{IPRW}(s,k)=\frac{\prod_{j=1}^{k}\Pr(Q_{s+30j-30}=0|\bar A_{s+30j-30}=0,\bar Q_{s+30j-60}=0,L_s,Z)}{\prod_{j=1}^{k}\Pr(Q_{s+30j-30}=0|\bar A_{s+30j-30}=0,\bar Q_{s+30j-60}=0,L_{s+30j-30},Z)}.
\end{equation}
The total weight for a given individual in period $k$  after the landmark is the product $\mathrm{IPACW}(s,k)\times \mathrm{IPRW}(s,k)$. In period $k=0$ the weight is equal to 1 for all individuals in $D^{dev}_0$.
 
The probabilities used in the weights were estimated using logistic regression. The covariates included were the same as those listed above, with the exception that we included the MELD-NA score in $L_s$ instead of the individual components of the MELD-NA score (creatinine, bilirubin, INR, sodium). All continuous covariates were modelled using restricted cubic splines with 3 knots. The models also included current time as a covariate, which was modelled using a restricted cubic splines with four knots placed at $30k$ corresponding to $180, 360, 720, 900$ days.

A Cox model was fitted using the dataset $D^{dev,cens}_0$, with landmark time as time zero. The model was fitted using the time-dependent weights ($\mathrm{IPACW}(s,k)\times \mathrm{IPRW}(s,k)$: updated every 30 days) and using predictors $Z$ and $L_s$ as measured at the landmark time. All continuous covariates were modelled using restricted cubic splines with 3 knots. The models also included $s$ (the landmark time) as a covariate, which was modelled using a restricted cubic splines with four knots placed at $s$ corresponding to $180, 360, 720, 900$ days.

\subsection{Validation}
\label{subsec:application.suppl.val}

The validation dataset $D^{val}=\{D_0^{val},D_1^{val}\}$ includes individuals for whom a prediction under the two interventions could be made at a range of times $s$. We applied the proposed approach of artificial censoring to the dataset $D^{val}$ to generate validation datasets $V^1$ and $V^0$ mimicking the two strategies. To mimic the strategy of receiving a transplant at a time $s$ in $V^1$, person-landmark observations in $D_0^{val}$ are artificially censored immediately at the landmark time, whereas follow-up on individuals in $D_1^{val}$ is retained. To mimic the strategy of not receiving a transplant at a time $s$ or in the future in $V^0$, individuals in $D_1^{val}$ are artificially censored immediately at time zero (the time of transplant), and person-landmark observations in $D_0^{val}$ are artificially censored at the time of transplant, if they received a transplant.

Created in this way, $V^1$ contains individuals artificially censored at time zero (the landmark time) but there is no artificial censoring at later times. We therefore estimate time-fixed IPACW, which are the inverse of the probability of receiving a transplant at time 0. The weight for an individual in $V^1$ transplanted at time $s$ is $G^{-1}_{1}(t|L,Z)=1/\Pr(A_{s}=1|\bar A_{s-1}=0,L_{s},Z)$, where $t$ denotes follow-up time starting from transplant. These weights are the same at all times $t$.

In $V^0$, individuals 
have been artificially censored due to transplant at a range of times. We therefore require time-dependent IPACW. Censoring due to improvement-based-removal from the waitlist also needs to be accounted for. The IPACW and IPRW were estimated using a similar approach as in the development step, after dividing each individual’s follow-up into periods of length 30 days, albeit using unstabilised rather than stabilised weights. For landmark time $s$ the total weight at times $30k\leq t<30(k+1)$ after the landmark time is 
 \begin{equation}
 \begin{split}
 G^{-1}_{0}(t|L,Z)&=\frac{1}{\prod_{j=1}^{k}\Pr(A_{s+30j}=0|\bar A_{s+30j-30}=0,Q_{s+30j-30}=0,L_{s+30j},Z)} \times \\
 & \quad \frac{1}{\prod_{j=1}^{k}\Pr(Q_{s+30j-30}=0|\bar A_{s+30j-30}=0,\bar Q_{s+30j-60}=0,L_{s+30j-30},Z)}.
 \end{split}
 \end{equation}
 Unlike in the development step, in the validation data the weights are not equal to 1 in period $k=0$ (i.e. at follow-up times $0\leq t<30$), because the weights also account for censoring of individuals 
 at time 0 if they are transplanted at that time. In particular the IPACW weights (first term in $G^{-1}_{0}(t|L,Z)$) are not equal to 1 in period $k=0$, whereas the IPRW (second term in $G^{-1}_{0}(t|L,Z)$) are still equal to 1 in period $k=0$. 

 In addition to the artificial censoring and the censoring due to improvement-based-removal, there is administrative censoring due to the end of follow-up at 3 years or at 30 April 2019. This additional standard censoring needs to be addressed in some measures of counterfactual predictive performance using a further set of weights, $G_{c}^{-1}(t)$, as described in the main manuscript. We used the Kaplan-Meier estimator on the observed validation data $D^{val}$ (so before artificial censoring was applied) to estimate the probability of remaining administratively uncensored up to any given time, where time is measured relative to the time of transplant (for individuals in $D_1^{val}$) or relative to the landmark time (for person-landmark observations in $D_0^{val}$). The weights $G_{c}^{-1}(t)$ are given by the inverse of these probabilities.

The models developed to enable predictions under the two intervention strategies of interest were used to obtain estimated risks under both strategies for each person-landmark observation in $D^{val}$ using their characteristics at time zero, which is the time of transplant for individuals in $D_1^{val}$ and the landmark time for person-landmark observations in $D_0^{val}$. After creating the validation datasets $V^1$ and $V^0$ and estimating the weights $G^{-1}_{a}(t|L,Z)$ and $G_{c}^{-1}(t)$ we obtained estimates of the measures of counterfactual predictive performance using the methods described in the main text.

We also applied the subset approach. For this, the predictions for each person-landmark observation in $D^{val}$ under the intervention strategy of receiving a transplant now were evaluated on the subset of individuals who received a transplant based on their post-transplant follow-up, i.e. on $D_1^{val}$. The predictions for each person-landmark observation in $D^{val}$ under the intervention strategy of not receiving a transplant now or in the future were evaluated on the subset of individuals who never received a transplant, from their landmark times onwards, i.e. on the subset of $D_0^{val}$ that never received a transplant. Some individuals in the subset of $D_0^{val}$ that never received a transplant are censored due to improvement-based removal from the waitlist. In the subset validation approach this censoring was treated in the same way as administrative censoring. 

\clearpage

\begin{table}[]
\caption{Simulation study: Data generating mechanisms and weight specifications, additive hazards model}\label{tab:sim.datagen}
    \centering
    (a) Models for $A_k$ and $L_k$: same in development and validation data unless stated otherwise, where $N(\mu,\sigma^2)$ denotes a normal distribution with mean $\mu$ and variance $\sigma^2$.
    \begin{tabular}{ll}
\hline
All scenarios &  $U\sim N(0,2^2)$\\
 & $L_0\sim N(10+U,4^2)$\\
 & $\mathrm{logit} \Pr(A_0=1|L_0)=-2+0.1 L_0$\\
 & $L_k \sim N(0.8 L_{k-1}- A_{k-1}+0.1 k+U,4^2)$ ($k\geq 1$)\\
 & $A_k=1$ if $A_{k-1}=1$, ($k\geq 1$)\\
 & $\mathrm{logit} \Pr(A_k=1|\bar A_{k-1},\bar L_{k})=\gamma_0+\gamma_L L_k$ if $A_{k-1}=0$, ($k\geq 1$)\\
 & with\\
 & $\gamma_0=-2$ and $\gamma_L=0.1$\\
Scenario 3 &  measurement error in risk calculation: $L_0^*=L_0+\epsilon,\quad \epsilon\sim N(0,16)$\\
Scenario 4a & positivity violation validation data: $\gamma_0=-0.25$\\
Scenario 4b & positivity violation validation data: $\gamma_0=-0.75$\\
Scenario 6c & quadratic term in model for $A_k$ validation data:\\
            & $\mathrm{logit} \Pr(A_0=1|L_0)= -1 + 0.01 L_0 + 0.01 L^2_0$\\
            & $\mathrm{logit} \Pr(A_k=1|\bar A_{k-1},\bar L_{k}) = -1 + 0.01 L_k + 0.01 L^2_k$ if $A_{k-1}=0$, ($k\geq 1$)\\
\hline
\end{tabular}
(b) Models for the conditional hazard $h(k|\bar A_{\lfloor k \rfloor},\bar{L}_{\lfloor k \rfloor},U)$
\begin{tabular}{lll}
\hline
 &Development data &Validation data\\
\hline
Scenario 1&$\alpha_{0}+\alpha_{A}A_{\lfloor k \rfloor}+\alpha_{L}(1-0.2(k-1))L_{\lfloor k \rfloor}+\alpha_U U$ & Same as development data\\
&with&\\
&$\alpha_{0}=0.2,\alpha_{A}=-0.04,\alpha_{L}=0.01,\alpha_U=0.01$&\\
Scenario 2&As in Scenario 1 but with $\alpha_{0}=0.3$& As in Scenario 1\\
Scenarios 3-6&As in Scenario 1& As in Scenario 1\\
    \hline
    \end{tabular}
(c) Weight models for $A_k, A_{k-1}=0$    
\begin{tabular}{llrcl}
\hline
Scenarios 1-4   & correctly specified & $\mathrm{logit} \Pr(A_k=1|\bar A_{k-1},\bar L_{k})$   & $\sim$ &   $1 + L_k$  \\
Scenario 5a     & exchangeability violation &             & $\sim$ &   $1$ \\
Scenario 5b     & exchangeability violation &             & $\sim$ &   $1 + L_0$ \\
Scenario 6a     & misspecified  &                         & $\sim$ &   $1 + \log(L_k+20)$ \\
Scenario 6b     & misspecified &                          & $\sim$ &   $1 + L^2_k$ \\
Scenario 6c     & misspecified &                          & $\sim$ &   $1 + L_k$  \\
Scenario 6d     & misspecified & $\mathrm{cauchit} \Pr(A_k=1|\bar A_{k-1},\bar L_{k})$   & $\sim$ &   $1 + L_k$  \\
\hline
\end{tabular}
\end{table}

\begin{table}[]
\caption{Simulation study: Data generating mechanisms Cox model}\label{tab:sim.datagen.cox}
    \centering
    (a) Models for $A_k$ and $L_k$: same in development and validation data unless stated otherwise, where $N(\mu,\sigma^2)$ denotes a normal distribution with mean $\mu$ and variance $\sigma^2$.
\begin{tabular}{ll}
\hline
All scenarios & $U\sim N(0,0.1^2)$\\
              & $L_0\sim N(U,1)$\\
              & $\mathrm{logit} \Pr(A_0=1|L_0)=\gamma_0+\gamma_L L_0$\\
              & $L_k \sim N(0.8 L_{k-1}- A_{k-1}+0.1 k+U,1^2)$ ($k\geq 1$)\\
              & $A_k=1$ if $A_{k-1}=1$, ($k\geq 1$)\\
              & $\mathrm{logit} \Pr(A_k=1|\bar A_{k-1},\bar L_{k})=\gamma_0+\gamma_L L_k$ if $A_{k-1}=0$, ($k\geq 1$)\\
              & with\\
              & $\gamma_0=-1, \gamma_L=0.5$\\
Scenario 3 &  measurement error in risk calculation: $L_0^*=L_0+\epsilon,\quad \epsilon\sim N(0,1)$\\
Scenario 4a & positivity violation validation data: $\gamma_0=0.5$\\
Scenario 4b & positivity violation validation data: $\gamma_0=0$\\
Scenario 6c & quadratic term in model for $A_k$ validation data:\\
            & $\mathrm{logit} \Pr(A_0=1|L_0)= -1 + 0.5 L_0 + 0.25 L^2_0$\\
            & $\mathrm{logit} \Pr(A_k=1|\bar A_{k-1},\bar L_{k}) = -1 + 0.5 L_k + 0.25 L^2_k$ if $A_{k-1}=0$, ($k\geq 1$)\\
\hline
\end{tabular}

(b) Models for the log of the conditional hazard: $\log(h(k|\bar A_{\lfloor k \rfloor},\bar{L}_{\lfloor k \rfloor}))$
\begin{tabular}{lll}
\hline
 &Development data & Validation data\\
\hline
Scenario 1&$\alpha_{0}+\alpha_{A}A_{\lfloor k \rfloor}+\alpha_{L}L_{\lfloor k \rfloor}+\alpha_U U$ & Same as development data\\
&with&\\
&$\alpha_{0}=-2,\alpha_{A}=-0.5,\alpha_{L}=0.5,\alpha_U=0.5$&\\
Scenario 2&As in Scenario 1 but with $\alpha_{0}=-1$& As in Scenario 1\\
Scenarios 3-6&As in Scenario 1& As in Scenario 1\\
    \hline
    \end{tabular}

(c) Weight models for $A_k, A_{k-1}=0$    
\begin{tabular}{llrcl}
\hline
Scenarios 1-4   & correctly specified & $\mathrm{logit} \Pr(A_k=1|\bar A_{k-1},\bar L_{k})$   & $\sim$ &   $1 + L_k$  \\
Scenario 5a     & exchangeability violation &             & $\sim$ &   $1$ \\
Scenario 5b     & exchangeability violation &             & $\sim$ &   $1 + L_0$ \\
Scenario 6a     & misspecified  &                         & $\sim$ &   $1 + \log(L_k+20)$ \\
Scenario 6b     & misspecified &                          & $\sim$ &   $1 + L^2_k$ \\
Scenario 6c     & misspecified &                          & $\sim$ &   $1 + L_k$  \\
Scenario 6d     & misspecified & $\mathrm{cauchit} \Pr(A_k=1|\bar A_{k-1},\bar L_{k})$   & $\sim$ &   $1 + L_k$  \\

\hline
\end{tabular}
\end{table}

\begin{table}[]
    \caption{Summary of counterfactual performance measures for assessment of predictions under interventions}\label{tab:estimands}
    {\small
    \centering
    \begin{tabular}{p{10cm}l}
    \hline
    Description & Notation\\
    \hline
    \multicolumn{2}{l}{{\bf Calibration}}\\
    Mean estimated risk by time $\tau$ under strategy $\underline{a}_0$   & $\bar{R}^{\underline{a}_0}(\tau)$,  $a=\mathbf{0},\mathbf{1}$ \\
    Counterfactual outcome proportion by time $\tau$ under strategy $\underline{a}_0$       & $\bar{R}_{\mathrm{Obs}}^{\underline{a}_0}(\tau)$,  $a=\mathbf{0},\mathbf{1}$\\ 
    Ratio of ``observed'' versus ``expected'' risk by time $\tau$ under strategy $\underline{a}_0$ & $\bar{R}_{\mathrm{Obs}}^{\underline{a}_0}(\tau)/\bar{R}^{\underline{a}_0}(\tau)$, $a=\mathbf{0},\mathbf{1}$\\
    
    &\\
    Mean estimated risks $\bar{R}^{\underline{a}_0}(\tau)$ within tenths of the estimated risks by time $\tau$ under strategy $\underline{a}_0$&-\\
    Counterfactual outcome proportions within tenths of the estimated risks by time $\tau$ under strategy $\underline{a}_0$&-\\
    &\\
    \multicolumn{2}{l}{{\bf Discrimination}}\\
      C-index truncated at time $\tau$ under strategy $\underline{a}_0$ & ${C}^{\underline{a}_0}(\tau)$,  $a=\mathbf{0},\mathbf{1}$ \\
      Cumulative/dynamic area under the ROC curve at time $t$ under strategy $\underline{a}_0$ & ${AUC}^{\underline{a}_0}_{C/D}(t)$,  $a=\mathbf{0},\mathbf{1}$ \\
      &\\
    \multicolumn{2}{l}{{\bf Brier score}}\\
    Brier score at time $t$ under strategy $\underline{a}_0$ & $BS^{\underline{a}_0}(t)$, $a=\mathbf{0},\mathbf{1}$ \\
    Scaled Brier score at time $t$ under strategy $\underline{a}_0$ & 
    $1-BS^{\underline{a}_0}(t) / BS_{0}^{\underline{a}_0}(t)$, $a=\mathbf{0},\mathbf{1}$ \\
      &\\      
            \hline
    \end{tabular}}
\end{table}

\begin{table}[ht]
\caption{Simulation results, Scenario 1, additive hazards model. Performance measures were obtained from the perfect validation data (true) and estimated from the observational validation data using the subset approach (subset) and using the proposed artificial censoring + inverse probability weighting estimators for assessing counterfactual performance (counterfactual). Results are averaged over 1000 simulation runs of validation datasets of size 3000.}
\label{tab:addhaz_sc1}
\begin{tabular}{lrllrll}
 \hline
 & \multicolumn{3}{c}{never treated} & \multicolumn{3}{c}{always treated} \\
 & true & subset & counterfactual & true & subset & counterfactual\\
\cmidrule(l){2-4} \cmidrule(l){5-7}
  
  \multicolumn{7}{l}{\bf Calibration: OE ratio based on risk by time 5}\\
mean & 1.002 & 1.145 & 1.002 & 1.003 & 1.003 & 1.003 \\ 
bias (SE) &  & 0.143 (0.001) & -0.000 (0.001) &  & 0.000 (0.001) & 0.000 (0.001) \\ 
  &\\
    \multicolumn{7}{l}{\bf Discrimination: C-index up to time 5}\\
mean & 0.546 & 0.578 & 0.547 & 0.555 & 0.552 & 0.556 \\ 
bias (SE) &  & 0.031 (0.000) & 0.000 (0.000) &  & -0.003 (0.000) & 0.001 (0.000) \\ 
&\\
  \multicolumn{7}{l}{\bf Discrimination: AUCt at time 5}\\
mean & 0.571 & 0.629 & 0.572 & 0.580 & 0.578 & 0.582 \\ 
  bias (SE) &  & 0.058 (0.001) & 0.000 (0.001) &  & -0.003 (0.001) & 0.001 (0.001) \\ 
 &\\
   \multicolumn{7}{l}{\bf Prediction error: scaled Brier score (\%) at time 5}\\
mean & 1.201 & -3.036 & 1.177 & 1.723 & 1.474 & 1.655 \\ 
bias (SE) &  & -4.237 (0.076) & -0.024 (0.033) &  & -0.250 (0.033) & -0.068 (0.042) \\ 
 \hline
 \multicolumn{7}{l}{\small OE ratio: ratio of observed versus expected risk.}\\
 \multicolumn{7}{l}{\small AUCt: cumulative/dynamic area under the ROC curve.}
 \end{tabular}
\end{table}

\begin{table}[ht]
\caption{Simulation results, Scenario 2, additive hazards model. Performance measures were obtained from the perfect validation data (true) and estimated from the observational validation data using the subset approach (subset) and using the proposed artificial censoring + inverse probability weighting estimators for assessing counterfactual performance (counterfactual). Results are averaged over 1000 simulation runs of validation datasets of size 3000.}
\label{tab:addhaz_sc2}
\begin{tabular}{lrllrll}
 \hline
 & \multicolumn{3}{c}{never treated} & \multicolumn{3}{c}{always treated} \\
 & true & subset & counterfactual & true & subset & counterfactual\\
\cmidrule(l){2-4} \cmidrule(l){5-7}
  
  \multicolumn{7}{l}{\bf Calibration: OE ratio based on risk by time 5}\\
mean & 0.858 & 0.973 & 0.857 & 0.809 & 0.822 & 0.809 \\ 
  bias (SE) &  & 0.115 (0.001) & -0.000 (0.001) &  & 0.014 (0.001) & -0.000 (0.001) \\&\\
    \multicolumn{7}{l}{\bf Discrimination: C-index up to time 5}\\
mean & 0.546 & 0.578 & 0.547 & 0.555 & 0.552 & 0.556 \\ 
  bias (SE) &  & 0.031 (0.000) & 0.000 (0.000) &  & -0.003 (0.000) & 0.001 (0.000) \\
  &\\
  \multicolumn{7}{l}{\bf Discrimination: AUCt at time 5}\\
mean & 0.571 & 0.629 & 0.572 & 0.580 & 0.578 & 0.582 \\ 
  bias (SE) &  & 0.058 (0.001) & 0.000 (0.001) &  & -0.003 (0.001) & 0.001 (0.001) \\
  &\\
   \multicolumn{7}{l}{\bf Prediction error: scaled Brier score (\%) at time 5}\\
mean & -5.416 & 1.932 & -5.421 & -7.736 & -7.149 & -7.802 \\ 
  bias (SE) &  & 7.347 (0.024) & -0.006 (0.080) &  & 0.587 (0.082) & -0.066 (0.089) \\

 \hline
  \multicolumn{7}{l}{\small OE ratio: ratio of observed versus expected risk.}\\
 \multicolumn{7}{l}{\small AUCt: cumulative/dynamic area under the ROC curve.}
\end{tabular}
\end{table}

\begin{table}[ht]
\caption{Simulation results, Scenario 3, additive hazards model. Performance measures were obtained from the perfect validation data (true) and estimated from the observational validation data using the subset approach (subset) and using the proposed artificial censoring + inverse probability weighting estimators for assessing counterfactual performance (counterfactual). Results are averaged over 1000 simulation runs of validation datasets of size 3000.}
\label{tab:addhaz_sc3}
\begin{tabular}{lrllrll}
 \hline
 & \multicolumn{3}{c}{never treated} & \multicolumn{3}{c}{always treated} \\
 & true & subset & counterfactual & true & subset & counterfactual\\
\cmidrule(l){2-4} \cmidrule(l){5-7}
    \multicolumn{7}{l}{\bf Calibration: OE ratio based on risk by time 5}\\
mean & 1.008 & 1.152 & 1.008 & 1.011 & 1.010 & 1.011 \\ 
  bias (SE) &  & 0.145 (0.001) & -0.000 (0.001) &  & -0.000 (0.001) & 0.000 (0.001) \\     
  &\\
    \multicolumn{7}{l}{\bf Discrimination: C-index up to time 5}\\
mean & 0.535 & 0.557 & 0.535 & 0.541 & 0.538 & 0.542 \\ 
  bias (SE) &  & 0.023 (0.000) & 0.000 (0.000) &  & -0.003 (0.000) & 0.001 (0.000) \\ 
  &\\
  \multicolumn{7}{l}{\bf Discrimination: AUCt at time 5}\\
mean & 0.554 & 0.596 & 0.554 & 0.560 & 0.557 & 0.561 \\ 
  bias (SE) &  & 0.042 (0.001) & 0.000 (0.001) &  & -0.003 (0.001) & 0.001 (0.001) \\   &\\
   \multicolumn{7}{l}{\bf Prediction error: scaled Brier score (\%) at time 5}\\
mean & -0.001 & -5.159 & -0.030 & -0.045 & -0.125 & -0.099 \\ 
  bias (SE) &  & -5.158 (0.088) & -0.029 (0.043) &  & -0.080 (0.048) & -0.054 (0.060) \\ 
 \hline
  \multicolumn{7}{l}{\small OE ratio: ratio of observed versus expected risk.}\\
 \multicolumn{7}{l}{\small AUCt: cumulative/dynamic area under the ROC curve.}
\end{tabular}
\end{table}

\begin{table}[ht]
\caption{Simulation results, Scenario 1, Cox model. Performance measures were obtained from the perfect validation data (true) and estimated from the observational validation data using the subset approach (subset) and using the proposed artificial censoring + inverse probability weighting estimators for assessing counterfactual performance (counterfactual). Results are averaged over 1000 simulation runs of validation datasets of size 3000.}
\label{tab:cox.sc1}
\begin{tabular}{lrllrll}
 \hline
 & \multicolumn{3}{c}{never treated} & \multicolumn{3}{c}{always treated} \\
 & true & subset & counterfactual & true & subset & counterfactual\\
\cmidrule(l){2-4} \cmidrule(l){5-7}
  
  \multicolumn{7}{l}{\bf Calibration - OE ratio based on risk by time 5}\\
mean & 0.989 & 1.217 & 0.986 & 1.004 & 1.006 & 1.004 \\ 
  bias (SE) &  & 0.227 (0.001) & -0.003 (0.002) &  & 0.002 (0.002) & 0.001 (0.003) \\ 
  &\\
    \multicolumn{7}{l}{\bf Discrimination: C-index up to time 5}\\
mean & 0.600 & 0.626 & 0.600 & 0.608 & 0.608 & 0.609 \\ 
  bias (SE) &  & 0.026 (0.000) & 0.000 (0.000) &  & -0.001 (0.001) & 0.001 (0.001) \\
  &\\
  \multicolumn{7}{l}{\bf Discrimination: AUCt at time 5}\\
mean & 0.626 & 0.675 & 0.626 & 0.618 & 0.619 & 0.619 \\ 
  bias (SE) &  & 0.050 (0.001) & 0.001 (0.001) &  & 0.001 (0.001) & 0.001 (0.001) \\
  &\\
   \multicolumn{7}{l}{\bf Prediction error: scaled Brier score (\%) at time 5}\\

mean & 4.312 & 0.007 & 4.326 & 3.237 & 3.443 & 3.224 \\ 
  bias (SE) &  & -4.305 (0.110) & 0.014 (0.081) &  & 0.206 (0.041) & -0.013 (0.041) \\

 \hline
  \multicolumn{7}{l}{\small OE ratio: ratio of observed versus expected risk.}\\
 \multicolumn{7}{l}{\small AUCt: cumulative/dynamic area under the ROC curve.}

\end{tabular}
\end{table}

\begin{table}[ht]
\caption{Simulation results, Scenario 2, Cox model. Performance measures were obtained from the perfect validation data (true) and estimated from the observational validation data using the subset approach (subset) and using the proposed artificial censoring + inverse probability weighting estimators for assessing counterfactual performance (counterfactual). Results are averaged over 1000 simulation runs of validation datasets of size 3000.}
\label{tab:cox.sc2}
\begin{tabular}{lrllrll}
 \hline
 & \multicolumn{3}{c}{never treated} & \multicolumn{3}{c}{always treated} \\
 & true & subset & counterfactual & true & subset & counterfactual\\
\cmidrule(l){2-4} \cmidrule(l){5-7}
  
  \multicolumn{7}{l}{\bf Calibration - OE ratio based on risk by time 5}\\
mean & 0.675 & 0.818 & 0.673 & 0.480 & 0.490 & 0.480 \\ 
  bias (SE) &  & 0.143 (0.001) & -0.002 (0.001) &  & 0.011 (0.001) & 0.000 (0.001) \\
  &\\
     \multicolumn{7}{l}{\bf Discrimination: C-index up to time 5}\\
mean & 0.600 & 0.626 & 0.600 & 0.608 & 0.608 & 0.609 \\ 
  bias (SE) &  & 0.026 (0.000) & 0.000 (0.000) &  & -0.001 (0.001) & 0.001 (0.001) \\ 
  &\\
  \multicolumn{7}{l}{\bf Discrimination: AUCt at time 5}\\
mean & 0.626 & 0.675 & 0.626 & 0.618 & 0.619 & 0.619 \\ 
  bias (SE) &  & 0.050 (0.001) & 0.001 (0.001) &  & 0.001 (0.001) & 0.001 (0.001) \\
  &\\
   \multicolumn{7}{l}{\bf Prediction error: scaled Brier score (\%) at time 5}\\
mean & -29.033 & -4.608 & -29.228 & -36.333 & -38.210 & -36.50 \\ 
  bias (SE) &  & 24.425 (0.088) & -0.195 (0.190) &  & -1.877 (0.237) & -0.176 (0.230) \\ 
 \hline
\multicolumn{7}{l}{\small OE ratio: ratio of observed versus expected risk.}\\
\multicolumn{7}{l}{\small AUCt: cumulative/dynamic area under the ROC curve.}
\end{tabular}
\end{table}

\begin{table}[ht]
\caption{Simulation results, Scenario 3, Cox model. Performance measures were obtained from the perfect validation data (true) and estimated from the observational validation data using the subset approach (subset) and using the proposed artificial censoring + inverse probability weighting estimators for assessing counterfactual performance (counterfactual). Results are averaged over 1000 simulation runs of validation datasets of size 3000.}
\label{tab:cox.sc3}
\begin{tabular}{lrllrll}
 \hline
 & \multicolumn{3}{c}{never treated} & \multicolumn{3}{c}{always treated} \\
 & true & subset & counterfactual & true & subset & counterfactual\\
\cmidrule(l){2-4} \cmidrule(l){5-7}
  
  \multicolumn{7}{l}{\bf Calibration - OE ratio based on risk by time 5}\\
mean & 0.988 & 1.214 & 0.986 & 0.965 & 0.971 & 0.967 \\ 
  bias (SE) &  & 0.227 (0.001) & -0.002 (0.001) &  & 0.006 (0.002) & 0.001 (0.002) \\ 
  &\\
     \multicolumn{7}{l}{\bf Discrimination: C-index up to time 5}\\
mean & 0.571 & 0.588 & 0.570 & 0.577 & 0.575 & 0.577 \\ 
  bias (SE) &  & 0.018 (0.000) & -0.000 (0.000) &  & -0.001 (0.001) & 0.001 (0.001) \\ 
  &\\
  \multicolumn{7}{l}{\bf Discrimination: AUCt at time 5}\\
mean & 0.589 & 0.623 & 0.589 & 0.584 & 0.584 & 0.585 \\ 
  bias (SE) &  & 0.034 (0.001) & -0.000 (0.001) &  & 0.000 (0.001) & 0.001 (0.001) \\ 
  &\\
   \multicolumn{7}{l}{\bf Prediction error: scaled Brier score (\%) at time 5}\\
mean & -1.633 & -6.951 & -1.753 & -0.340 & -0.370 & -0.334 \\ 
  bias (SE) &  & -5.317 (0.126) & -0.120 (0.113) &  & -0.030 (0.063) & 0.006 (0.060) \\  
 \hline
 \multicolumn{7}{l}{\small OE ratio: ratio of observed versus expected risk.}\\
 \multicolumn{7}{l}{\small AUCt: cumulative/dynamic area under the ROC curve.}
\end{tabular}
\end{table}

\begin{sidewaystable}
\caption{Simulation results, Scenarios 4,5,6, additive hazards model. Bias (SE) of the performance measures obtained from the observational validation data using proposed artificial censoring + inverse probability weighting estimators for assessing counterfactual performance, compared to true performance obtained from the perfect validation data. Numbers displayed in bold indicate that results of the counterfactual performance measures had higher bias than those of the naive subset approach. Results are averaged over 1000 simulation runs of validation datasets of size 3000.}
\label{tab:addhaz_sc456}
\begin{tabular}{lllllllll}
 \hline
 & \multicolumn{4}{c}{never treated} & \multicolumn{4}{c}{always treated} \\
 scenario & OE ratio & C-index & AUCt & Brier (\%) & OE ratio & C-index & AUCt & Brier (\%)\\
  \cmidrule(l){2-5} \cmidrule(l){6-9}

 \multicolumn{9}{l}{4a: Positivity violation: on average only 31 individuals untreated after time point 4}\\
& -0.007 (0.003) & -0.002 (0.001) &	0.054 (0.004) &	3.421 (0.383) &	-0.001 (0.001) &	0.000 (0.000) &	0.001 (0.000) &	0.014 (0.024)\\

 \multicolumn{9}{l}{4b: Positivity violation: on average only 73 individuals untreated after time point 4} \\
& -0.003 (0.002) & -0.001 (0.001) & 0.021 (0.003) & 1.164 (0.169) & -0.000 (0.001) &	0.000 (0.000) &	0.000 (0.000) &	-0.010 (0.027)\\

 \multicolumn{9}{l}{5a: Exchangeability violation: weights do not use $L$}\\
& -0.032 (0.001) & -0.002 (0.000) & 0.041 (0.001) & 1.443 (0.037) & 0.034 (0.001) & -0.003 (0.000) & -0.003 (0.001) & -0.250 (0.033)\\

 \multicolumn{9}{l}{5b: Exchangeability violation: weights use $L_0$ instead of $L_k$}\\
& -0.013 (0.001) & -0.001 (0.000) & -0.030 (0.001) & -1.225 (0.048) &	0.000 (0.001) &	0.001 (0.000) &	0.001 (0.001) & -0.068 (0.042)\\

 \multicolumn{9}{l}{6a: Misspecified weights: weights use $\log(L_k)$ instead of $L_k$}\\
& -0.002 (0.001) & -0.001 (0.000) & 0.001 (0.001) & -0.022 (0.028) & 0.004 (0.001) &	 -0.001 (0.000) & -0.002 (0.001) & -0.195 (0.038) \\
 
 \multicolumn{9}{l}{6b: Misspecified weights: weights use $L^2_k$ instead of $L_k$}\\
& -0.012 (0.002) & 0.006 (0.001) & 0.058 (0.003) & 4.186 (0.277) & 0.006 (0.001) & -0.003 (0.000) & \textbf{-0.004 (0.001)} & \textbf{-0.305 (0.035)}\\

 \multicolumn{9}{l}{6c: Misspecified weights: $L^2_k$ in data generation, weights use $L_k$ only}\\

 & -0.014 (0.001) & -0.008 (0.000) & -0.015 (0.001) & 0.032 (0.033) & \textbf{-0.023 (0.002)} &	\textbf{0.013 (0.001) }&	\textbf{0.016 (0.001) }& \textbf{1.373 (0.092)}\\

 \multicolumn{9}{l}{6d: Misspecified weights: weights use cauchit link in stead of logit link}\\
& 0.001 (0.001) &	0.002 (0.000) &	0.003 (0.001) &	0.299 (0.068) &	0.000 (0.001) &	0.001 (0.000) & 0.001 (0.001) & -0.068 (0.042) \\

 \hline
\multicolumn{9}{l}{\small OE ratio: ratio of observed versus expected risk.}\\
\multicolumn{9}{l}{\small AUCt: cumulative/dynamic area under the ROC curve.}
\end{tabular}
\end{sidewaystable}

\begin{sidewaystable}
\caption{Simulation results, Scenarios 4,5,6, Cox model. Bias (SE) of the performance measures obtained from the observational validation data using proposed artificial censoring + inverse probability weighting estimators for assessing counterfactual performance, compared to true performance obtained from the perfect validation data. Numbers displayed in bold indicate that results of the counterfactual performance measures had higher bias than those of the naive subset approach. Results are averaged over 1000 simulation runs of validation datasets of size 3000.}
\label{tab:cox.sc456}
\begin{tabular}{lllllllll}
 \hline
 & \multicolumn{4}{c}{never treated} & \multicolumn{4}{c}{always treated} \\
 scenario & OE ratio & C-index & AUCt & Brier (\%) & OE ratio & C-index & AUCt & Brier (\%)\\
  \cmidrule(l){2-5} \cmidrule(l){6-9}

 \multicolumn{9}{l}{4a: Positivity violation: on average only 25 individuals untreated after time point 4}\\
    &   -0.026 (0.006) & -0.000 (0.002) & 0.034 (0.004) & 3.603 (0.720) & 0.000 (0.002) & 0.000 (0.000) & 0.000 (0.000) & 0.015 (0.027)\\

 \multicolumn{9}{l}{4b: Positivity violation: on average only 73 individuals untreated after time point 4} \\
    &   -0.009 (0.003) & -0.001 (0.001) & 0.007 (0.003) & 0.742 (0.294) & 0.001 (0.002) & 0.000 (0.001) & 0.000 (0.001) & 0.015 (0.031)\\

 \multicolumn{9}{l}{5a: Exchangeability violation: weights do not use $L$}\\
    &   -0.149 (0.001) & -0.006 (0.000) & 0.018 (0.001) & 0.832 (0.064) & \textbf{0.114 (0.003)} & -0.001 (0.001) & 0.001 (0.001) & 0.206 (0.041) \\

 \multicolumn{9}{l}{5b: Exchangeability violation: weights use $L_0$ instead of $L_k$}\\
    &   -0.108 (0.001) & -0.003 (0.000)	& -0.036 (0.001) &\textbf{ -4.972 (0.109)} & 0.001 (0.003) &	0.001 (0.001) &	0.001 (0.001) & -0.013 (0.041) \\

 \multicolumn{9}{l}{6a: Misspecified weights: weights use $\log(L_k)$ instead of $L_k$}\\
    &   -0.009 (0.001 & -0.001 (0.000) & -0.001 (0.001) & -0.096 (0.078) & -0.001 (0.003) & \textbf{0.002 (0.001)} &	\textbf{0.002 (0.001}) & 0.038 (0.041) \\

 \multicolumn{9}{l}{6b: Misspecified weights: weights use $L^2_k$ instead of $L_k$}\\
    &   -0.152 (0.001) & -0.004 (0.000) & 0.024 (0.001) & 1.212 (0.065) &\textbf{ 0.105 (0.003) }& \textbf{-0.004 (0.001)} & \textbf{-0.002 (0.001)} & -0.030 (0.038) \\

 \multicolumn{9}{l}{6c: Misspecified weights: $L^2_k$ in data generation, weights use $L_k$ only}\\
    &   -0.088 (0.002) & -0.022 (0.000) &\textbf{ -0.027 (0.001) }& -0.405 (0.087) &\textbf{	\textbf{-0.023} (0.002) }&\textbf{ 0.018 (0.001)} &\textbf{ 0.020 (0.001)} &\textbf{ 1.045 (0.046)} \\

 \multicolumn{9}{l}{6d: Misspecified weights: weights use cauchit link instead of logit link}\\
    &   0.008 (0.002 & 0.004 (0.000) & 0.006 (0.001) & 0.636 (0.096) & 0.001 (0.003) & 0.001 (0.001) & 0.001 (0.001) & -0.013 (0.041)  \\

 \hline
 \multicolumn{9}{l}{\small OE ratio: ratio of observed versus expected risk.}\\
 \multicolumn{9}{l}{\small AUCt: cumulative/dynamic area under the ROC curve.}
\end{tabular}
\end{sidewaystable}


\begin{table}[ht]
\centering
\caption{Liver transplant application, development data $D^{dev}$: Summary of covariates as recorded at time zero for the $n=116797$ person-landmark observations in the development data, where time zero is the time of transplant for person-landmark observations in $D_1^{dev}$ and the landmark time for person-landmark observations in $D_0^{dev}$. Summaries of numbers in the 11 regions are omitted. The numbers in each region ranged from 3660 (3.1\%) (region 6) to 18460 (15.8\%) (region 5).}
\label{tab:dev.desc}
\begin{subtable}[l]{\textwidth}
\caption{Time-fixed variables}
\footnotesize
\centering
\begin{tabular}{rll}
  \hline
 &  & n (\%) or Mean (SD) \\ 
  \hline
  Sex & Male & 78038 (66.8)  \\ 
  & Female &  38759 (33.2)  \\ 
  Ethnicity & Asian &  4227 ( 3.6)  \\ 
   & Black &  9303 ( 8.0)  \\ 
   & Other &  1999 ( 1.7)  \\  
   & White & 101268 (86.7)  \\
  Blood group & A & 44842 (38.4)  \\ 
   & AB &  3666 ( 3.1)  \\
   & B &  13848 (11.9)  \\
   & O & 54441 (46.6)  \\ 
   Disease group & Alcoholic Cirrhosis & 40360 (34.6)  \\ 
   & HBV Cirrhosis HBV &   2162 ( 1.9)  \\ 
   & HCV Cirrhosis HCV &  26975 (23.1)  \\ 
   & Cryptogenic &   383 ( 0.3)  \\  
   & HCC &  15022 (12.9)  \\ 
  & NASH &  23357 (20.0)  \\ 
   & PBC &  3428 ( 2.9)  \\ 
   & PSC &  5110 ( 4.4)  \\  
  Diabetes & No & 79116 (67.7)  \\ 
   & Yes &  37681 (32.3)  \\
  BMI & mean (SD) &  29.33 (5.86) \\
     \hline
\end{tabular}
\end{subtable}
\vspace{0.5cm}

\begin{subtable}[r]{\textwidth}
\caption{Time-dependent variables (as recorded at time zero)}
\footnotesize
\centering
\begin{tabular}{rll}
     \hline
 & & n (\%) or Mean (SD) \\ 
      \hline
   Age & mean (SD) & 58.90 (8.99) \\ 
  CKD & No &  78734 (67.4) \\ 
   & Yes & 38063 (32.6) \\ 
  Dialysis & No & 108797 (93.2) \\ 
   & Yes & 8000 (6.8) \\ 
  MELD-NA score & mean (SD) & 15.59 (9.22) \\ 
  Albumin& mean (SD) & 3.24 (0.66) \\ 
  Sodium & mean (SD) & 136.81 (4.44) \\ 
  Creatinine & mean (SD) &  1.32 (1.26) \\ 
  Bilirubin &  mean (SD) & 3.92 (6.29) \\ 
  INR & mean (SD) & 1.50 (0.64) \\ 
  Ascites & Absent & 37716 (32.3) \\ 
   & Slight & 55118 (47.2) \\ 
   & Moderate & 23963 (20.5) \\ 
 Encephalopathy & 1-2 & 59185 (50.7) \\ 
   & 3-4 & 6097 (5.2) \\ 
   & None & 51515 (44.1) \\ 
  Child Pugh Score grade & A & 26263 (22.5) \\ 
  & B & 51439 (44.0) \\ 
  & C & 39095 (33.5) \\ 
  Number of tumours & 0 & 89570 (76.7) \\  
   & 1 & 23942 (20.5) \\ 
   & $\geq$ 2 & 3285 (2.8) \\ 
  Exception points & No & 87172 (74.6) \\ 
   & Yes & 29625 (25.4) \\ 
  Exception points HCC & No & 105841 (90.6) \\ 
   & Yes & 10956 (9.4) \\  
   \hline
\end{tabular}
\end{subtable}
\end{table}

\begin{table}[ht]
\centering
\caption{Liver transplant application, validation data $D^{val}$: Summary of covariates as recorded at time zero for the $n=49966$ person-landmark observations in the validation data, where time zero is the time of transplant for person-landmark observations in $D_1^{val}$ and the landmark time for person-landmark observations in $D_0^{val}$. Summaries of numbers in the 11 regions are omitted. The numbers in each region ranged from 1529 (3.1\%) (region 6) to 7688 (15.4\%) (region 5).}
\label{tab:val.desc}
\begin{subtable}[l]{\textwidth}
\caption{Time-fixed variables}
\footnotesize
\centering
\begin{tabular}{rll}
  \hline
 &  & n (\%) or Mean (SD) \\ 
  \hline
 Sex & Male &  33529 (67.1)  \\ 
   & Female &  16437 (32.9)  \\ 
  Ethnicity & Asian &   1767 ( 3.5)  \\ 
  & Black &  3894 ( 7.8)  \\  
  & Other &  853 ( 1.7)  \\
  & White & 43452 (87.0)  \\ 
  Blood group & A & 19141 (38.3)  \\ 
   & AB & 1592 ( 3.2)  \\  
   & B &  5847 (11.7)  \\ 
   & O &  23386 (46.8)  \\ 
  Disease group & Alcoholic Cirrhosis & 17421 (34.9)  \\ 
   & HBV Cirrhosis & 931 ( 1.9)  \\  
   & HCV Cirrhosis &  11539 (23.1)  \\ 
  & Cryptogenic & 160 ( 0.3)  \\ 
   & HCC & 6225 (12.5)  \\ 
   & NASH &  10063 (20.1)  \\ 
   & PBC &  1426 ( 2.9)  \\ 
   & PSC &  2201 ( 4.4)  \\ 
 Diabetes & No &  33958 (68.0)  \\ 
   & Yes &  16008 (32.0)  \\ 
 BMI & mean (SD) & 29.32 (5.82) \\
     \hline
\end{tabular}
\end{subtable}
\vspace{0.5cm}

\begin{subtable}[r]{\textwidth}
\caption{Time-dependent variables (as recorded at time zero)}
\footnotesize
\centering
\begin{tabular}{rll}
     \hline
 & & n (\%) or Mean (SD) \\ 
      \hline
  Age & mean (SD) & 58.67 (9.11) \\ 
  CKD & No & 33603 (67.3) \\ 
   & Yes & 16363 (32.7) \\ 
  Dialysis & No & 46486 (93.0) \\ 
  & Yes & 3480 (7.0) \\
  MELD-NA score & mean (SD) & 15.62 (9.19) \\ 
  Albumin& mean (SD) & 3.25 (0.66) \\ 
  Sodium & mean (SD) & 136.79 (4.45) \\ 
  Creatinine & mean (SD) & 1.33 (1.28) \\ 
  Bilirubin &  mean (SD) & 3.88 (6.20) \\ 
  INR & mean (SD) & 1.50 (0.65) \\ 
  Ascites & Absent & 16079 (32.2) \\ 
   & Slight & 23485 (47.0) \\ 
   & Moderate & 10402 (20.8) \\ 
  Encephalopathy & 1-2 & 25377 (50.8) \\ 
   & 3-4 & 2627 (5.3) \\ 
   & None & 21962 (44.0) \\ 
  Child Pugh Score grade & A & 11176 (22.4) \\ 
   & B & 22161 (44.4) \\ 
   & C & 16629 (33.3) \\ 
  Number of tumours & 0 & 38351 (76.8) \\ 
   & 1 & 10200 (20.4) \\ 
   & $\geq$2 & 1415 (2.8) \\ 
  Exception points & 0 & 37283 (74.6) \\ 
   & 1 & 12683 (25.4) \\ 
  Exception points HCC & 0 & 45221 (90.5) \\ 
   & 1 & 4745 (9.5) \\
   \hline
\end{tabular}
\end{subtable}
\end{table}

\begin{table}[]
    \caption{Liver transplant application: Summary of numbers of events and censorings in the validation datasets created to mimic the ``no transplant'' and ``transplant'' strategies. }
    \label{tab:val.events}
    \centering
    \begin{tabular}{p{8cm}p{3cm}p{3cm}}
    \hline
         &  No transplant: $V^0$ & Transplant: $V^1$\\
    \hline
     Number of person-landmark observations    & 49966 & 49966\\
     Number of unique individuals    & 27929 & 27929\\
     Number of person-landmark observations artificially censored due to deviation from strategy&24439&42820\\
     Number of events (composite outcome) &7221 &777\\
     Number censored due to removal from waitlist due to improvement or for ``other'' reasons&7883&-\\
    Number administratively censored&10423&6369\\
     Total person-landmark-observation-years of follow-up&35478&13126\\
    \hline
    \end{tabular}
\end{table}

\begin{table}
\caption{Liver transplant application: Evaluation of model performance using the subset approach.}
\label{tab:liver:subset}
\begin{center}
\begin{tabular}{lcc}
\hline
&\multicolumn{2}{c}{Strategy}\\
\cline{2-3} 
&No transplant&Transplant\\
\hline
  Calibration: observed/expected ratio based on risk by 3 years  &1.058 &1.010\\
&&\\
  Discrimination: C-index up to 3 years&0.756&0.587\\
  Discrimination: AUCt at 3 years &0.783&0.578\\
  &&\\
  Prediction error: scaled Brier score (\%) at 3 years &18.2&0.46\\
  \hline
    {\small AUCt: cumulative/dynamic area under the receiver operating characteristic curve.}\\
\end{tabular}
\end{center}
\end{table}

\clearpage

\begin{figure}
	\centering
		\includegraphics[scale=0.5]{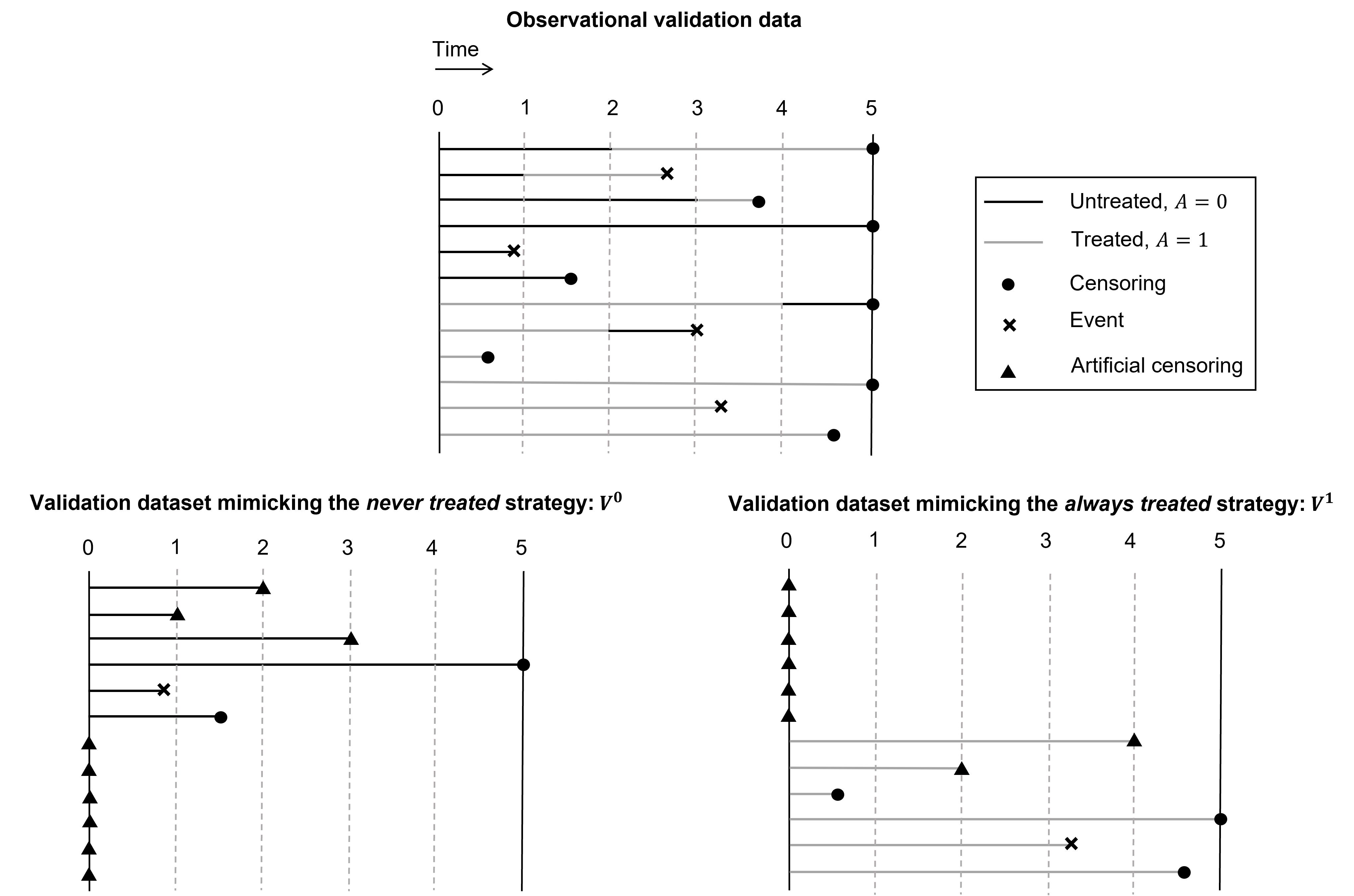}
			\caption{Illustration of how the validation datasets $V^{\underline{a}_0=0}$ and $V^{\underline{a}_0=1}$ are created from the observational validation data. The original data depicts observed follow-up for 12 individuals, with 5 visit times and administrative censoring at time 5. Treatment status $A_t$ is observed at times $t=0,1,2,3,4$ and assumed to be constant between visits. For the \emph{always treated} strategy, individuals are artificially censored at the time $t$ at which they have $A_t=0$, if this occurs before their end of follow-up. Individuals with $A_0=0$ are censored immediately at time 0, and those with $A_0=1$ who later switch to $A_t=0$ are censored at the time of the switch. For the \emph{never treated} strategy, individuals in the validation data are artificially censored when they deviate from the \emph{never treated} strategy, with some individuals (those with $A_0=1$) being censored at time 0.}\label{fig:method.suppl}
\end{figure}

\begin{figure}
	\centering
		\includegraphics[scale=1.0]{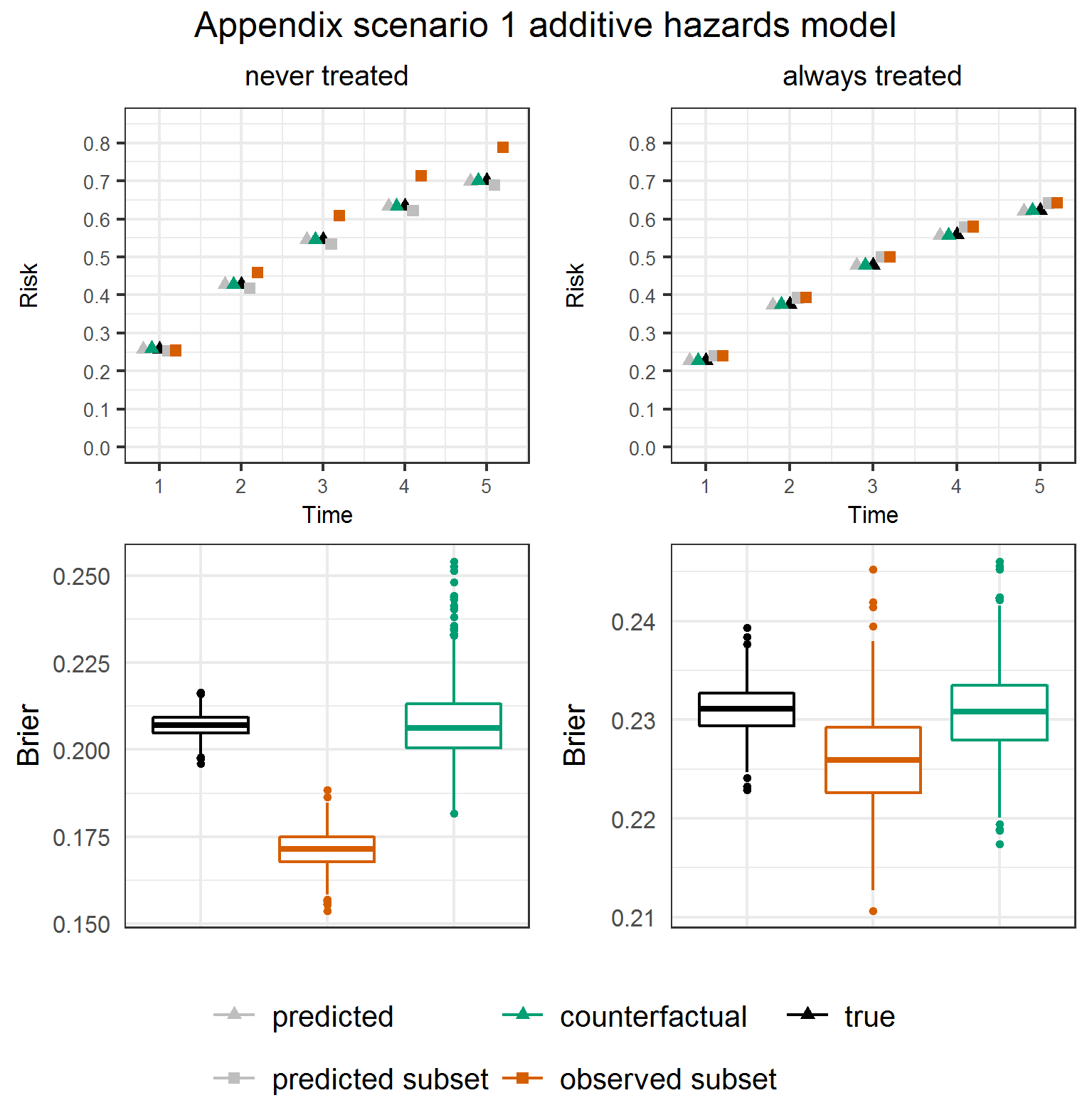}
	    \caption{Simulation results: additive hazards model Scenario 1. Left panel:  for the \emph{never treated} strategy. Right panel: for the \emph{always treated} strategy. Top row: outcome proportions over time estimated by the prediction model (grey triangles), observed in the perfect validaiton data (black triangles) and estimated from the observational validation data using the proposed artificial censoring + inverse probability weighting estimators for counterfactual performance assessment (green triangles). Estimated and observed outcome proportions using the subset approach are depicted with grey and orange squares.  Bottom row: unscaled Brier score at time 5.}\label{fig:sim.addhaz.scenario1.appendix}
\end{figure}

\begin{figure}
	\centering
		\includegraphics[scale=1.0]{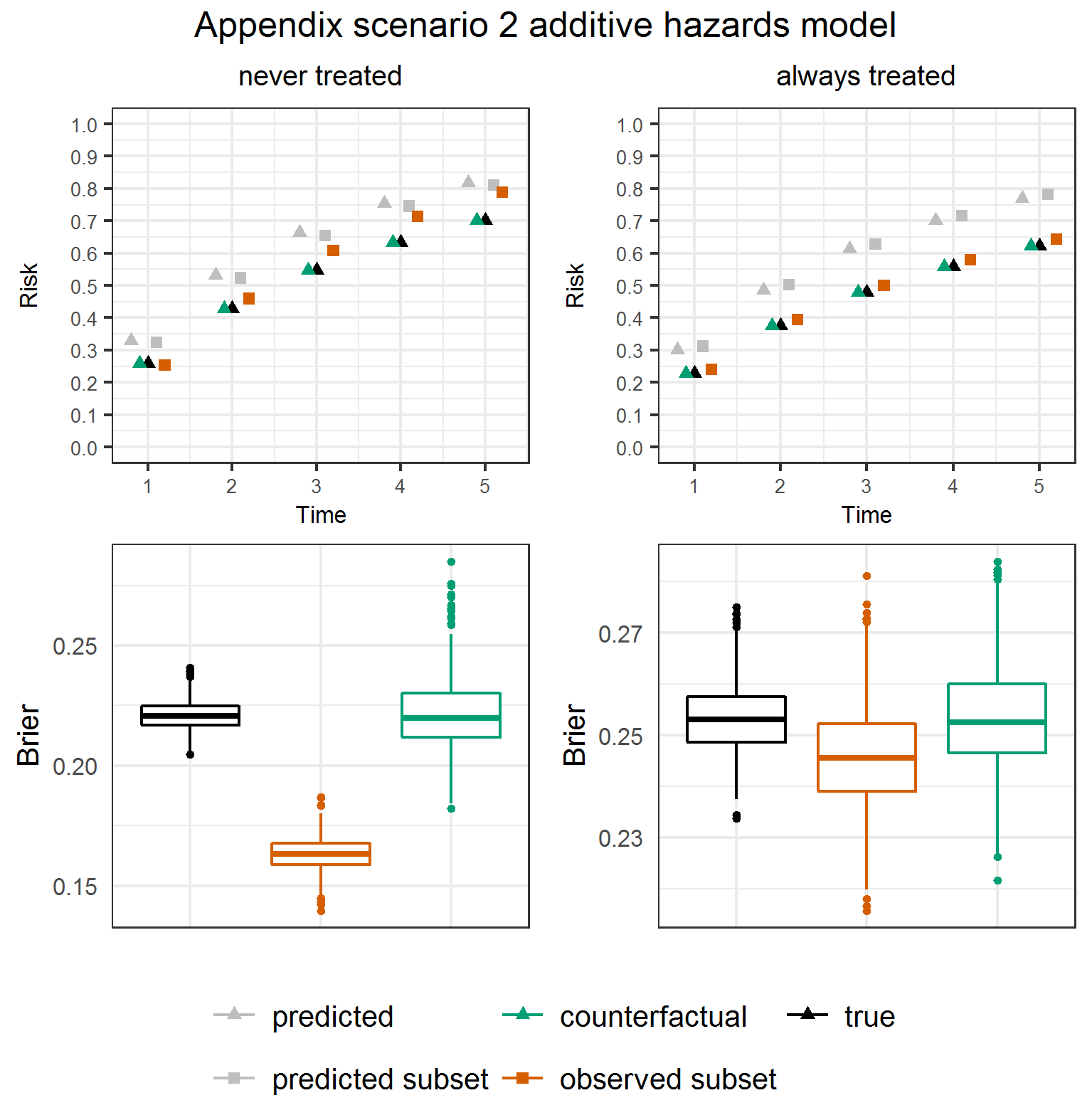}
	    \caption{Simulation results: additive hazards model Scenario 2. Left panel:  for the \emph{never treated} strategy. Right panel: for the \emph{always treated} strategy. Top row: outcome proportions over time estimated by the prediction model (grey triangles), observed in the perfect validaiton data (black triangles) and estimated from the observational validation data using the proposed artificial censoring + inverse probability weighting estimators for counterfactual performance assessment (green triangles). Estimated and observed outcome proportions using the subset approach are depicted with grey and orange squares.  Bottom row: unscaled Brier score at time 5. }\label{fig:sim.addhaz.scenario2.appendix}
\end{figure}

\begin{figure}
	\centering
		\includegraphics[scale=1.0]{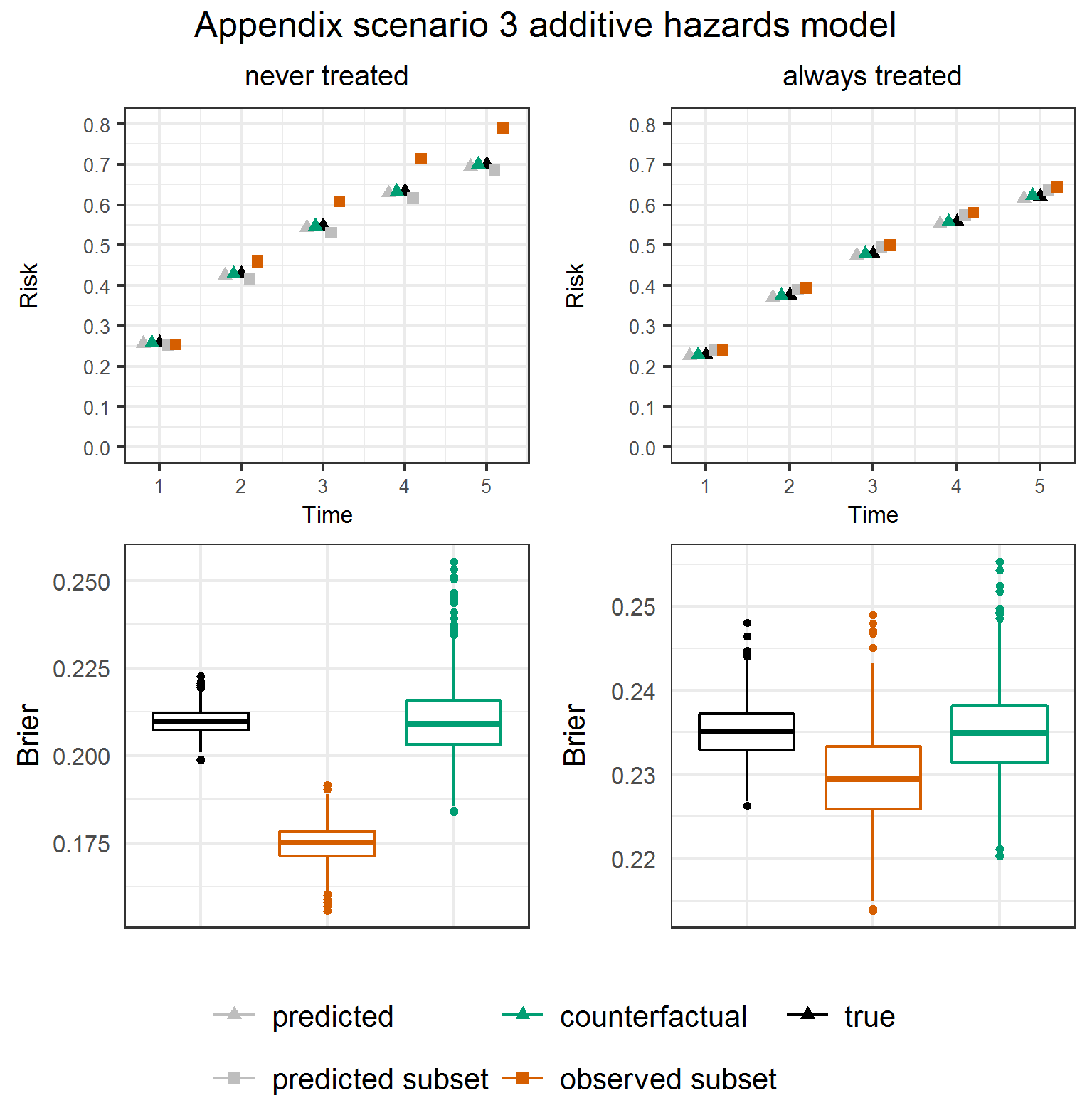}
	    \caption{Simulation results: additive hazards model Scenario 3. Left panel:  for the \emph{never treated} strategy. Right panel: for the \emph{always treated} strategy. Top row: outcome proportions over time estimated by the prediction model (grey triangles), observed in the perfect validaiton data (black triangles) and estimated from the observational validation data using the proposed artificial censoring + inverse probability weighting estimators for counterfactual performance assessment (green triangles). Estimated and observed outcome proportions using the subset approach are depicted with grey and orange squares.  Bottom row: unscaled Brier score at time 5. }\label{fig:sim.addhaz.scenario3.appendix}
\end{figure}

\begin{figure}
    \centering
    \includegraphics[scale=0.8]{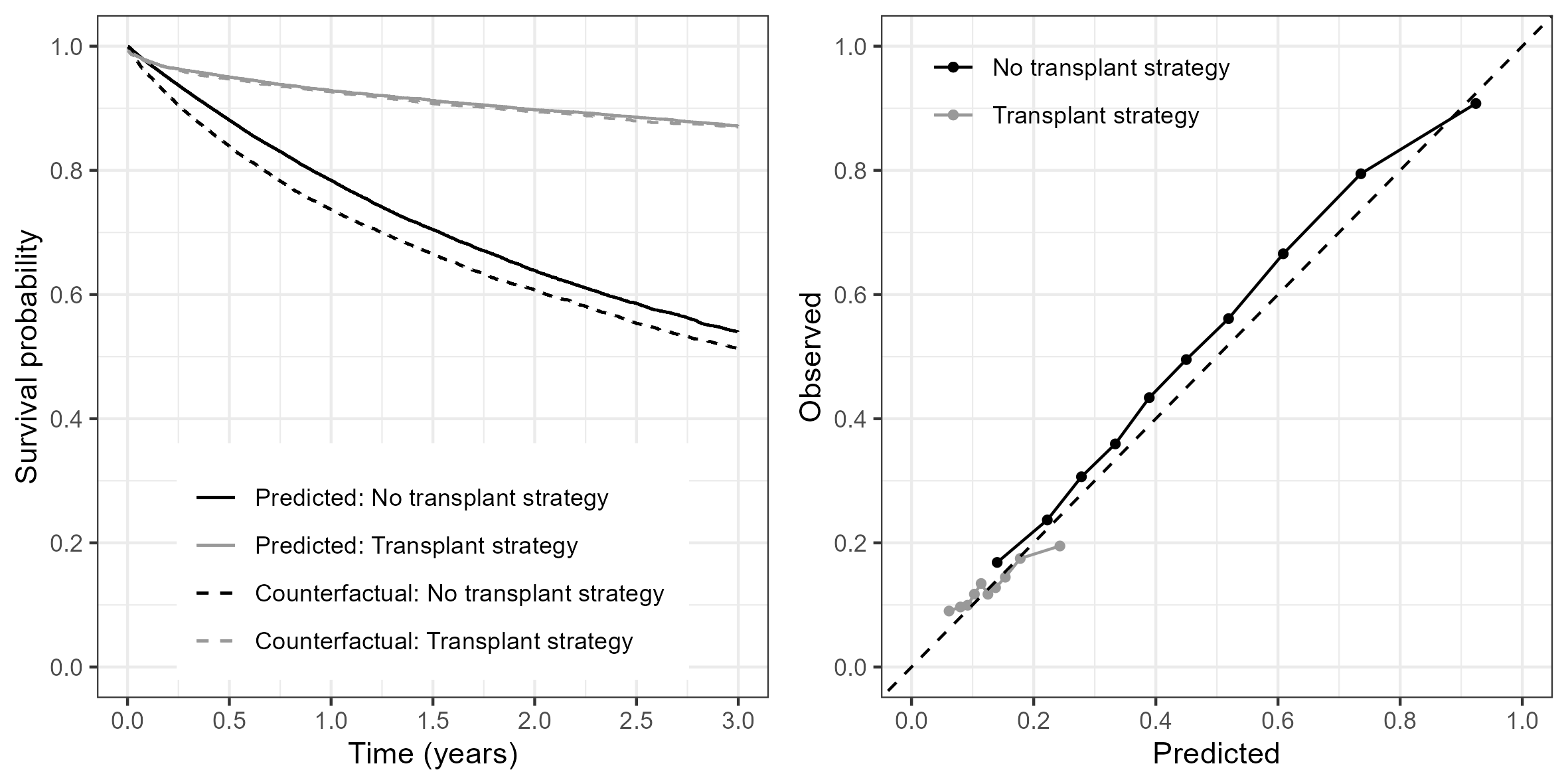}
    \caption{Liver transplant application: Calibration of estimated risks under the \emph{no transplant} and \emph{transplant} strategies. Left: Plot showing the mean estimated survival curves up to three years under the two transplant strategies (solid lines), and the corresponding observed survival curves (dashed lines), obtained using the subset approach. Right: Plot of mean observed outcome proportions by 3 years (obtained using the subset approach) against mean estimated risk by 3 years within 10 equal-sized groups of estimated risk under the two transplant strategies, showing the line of equality (dashed line).}
    \label{fig:liver.calibration.subset}
\end{figure}

\end{document}